\documentclass[12pt,preprint]{aastex}
\usepackage{epsfig}
\usepackage{amsmath}

\newcommand{\nH}{\ensuremath{\rm n_H}}

\newcommand{\degree}{{\rm o}}

\newcommand{\fcgs}{${\rm erg}\, {\rm cm}^{-2}\, {\rm s}^{-1}$}

\newcommand{\chisq}{$\chi^2$}

\newcommand{\Chandra}{{\sl Chandra}}

\newcommand{\Fermi}{{\sl Fermi}}

\newcommand{\Msun}{ \ensuremath{\rm M_\odot} }

\newcommand{\etal}{et~al.}
\newcommand{\FL}{PSR J2021+4026}

\newcommand{\gcyg}{$\gamma$~Cygni}
\newcommand{\gray}{{$\gamma$-ray}}
\newcommand{\grays}{{$\gamma$~rays}}

\begin{document}
\title{{The Identification of the X-ray Counterpart to PSR J2021+4026}}

\author{
 Martin~C.~Weisskopf\altaffilmark{1},
 Roger~W.~Romani\altaffilmark{2}, 
 Massimiliano~Razzano\altaffilmark{3,4,5},
 Andrea~Belfiore\altaffilmark{4,6,7},
 Pablo Saz Parkinson\altaffilmark{4},
 Paul~S.~Ray\altaffilmark{8},
 Matthew~Kerr\altaffilmark{9},
 Alice~Harding\altaffilmark{10},
 Douglas~A.~Swartz\altaffilmark{11},
 Alberto~Carrami\~nana\altaffilmark{12},
 Marcus~Ziegler\altaffilmark{4},
 Werner~Becker\altaffilmark{13}, 
Andrea~De~Luca\altaffilmark{6,14,15},
 Michael~Dormody\altaffilmark{4},
 David~J.~Thompson\altaffilmark{16},
 Gottfried~Kanbach\altaffilmark{13},
 Ronald~F.~Elsner\altaffilmark{1},
 Stephen~L.~O'Dell\altaffilmark{1},
 Allyn~F.~Tennant\altaffilmark{1}
}

\altaffiltext{1}
{NASA Marshall Space Flight Center, Space Science Office, VP62, Huntsville, AL 35812 USA.}
\altaffiltext{2}
{Department of Physics, Stanford University, Stanford, CA 94305}
\altaffiltext{3}
{Istituto Nazionale di Fisica Nucleare, Sezione di Pisa, I-56127 Pisa, Italy}
\altaffiltext{4} 
{Santa Cruz Institute for Particle Physics, Department of Physics,  University of California at Santa Cruz, Santa Cruz, CA 95064, USA}
\altaffiltext{5} 
{Dipartimento di Fisica, Università di Pisa, 56127 Pisa, Italy}
\altaffiltext{6} 
{INAF-Istituto di Astrofisica Spaziale e Fisica Cosmica, I-20133 Milano, Italy}
\altaffiltext{7} 
{Universit\'{a} di Pavia, Dipartimento di Fisica Teorica e Nucleare (DFNT), I-27100 Pavia, Italy}
\altaffiltext{8}
{Space Science Division, Naval Research Laboratory, Washington, DC 20375, USA }
\altaffiltext{9} 
{Kavli Institute for Particle Astrophysics and Cosmology, Stanford University, 452 Lomita Mall, Stanford, CA 94305, USA}
\altaffiltext{10}
{Astrophysics Science Division, NASA Goddard Space Flight Center,
Code 663, Greenbelt, MD 20771 USA.}
\altaffiltext{11}
{Universities Space Research Association, NASA Marshall Space Flight Center, Space Science Office, VP62, Huntsville, AL 35812 USA.}
\altaffiltext{12}
{Instituto Nacional de Astrof\'{\i}sica, \'{O}ptica y Electr\'{o}nica, Luis Enrique Erro 1, Tonantzintla, Puebla 72840, M\'{e}xico.}
\altaffiltext{13}
{Max-Planck-Institut f\"{u}r extraterrestrische Physik, 85741 Garching bei M\"{u}nchen, Germany.}
\altaffiltext{14}
{IUSS-Istituto Universitario di Studi Superiori, Viale Lungo Ticino Sforza 56, 27100 Pavia, Italy}
\altaffiltext{15}
{INFN-Istituto Nazionale di Fisica Nucleare, sezione di Pavia, via A. Bassi 6, 27100 Pavia, Italy}
\altaffiltext{16}
{Astroparticle Physics Laboratory, NASA Goddard Space Flight Center, Code 661, Greenbelt, MD 20771 USA.}
{  }

\begin{abstract}
We report the probable identification of the X-ray counterpart to the \gray\ pulsar PSR J2021+4026 using imaging with the Chandra X-ray Observatory ACIS and timing analysis with the \Fermi\ satellite.
Given the statistical and systematic errors, the positions determined by both satellites are coincident.
The X-ray source position is R.A. 20$^h$21$^m$30$^s$.733, Decl.\ $+$40\arcdeg26\arcmin46.04\arcsec\ (J2000)
 with an estimated uncertainty of 1.\arcsec3 combined statistical and systematic error.
Moreover, both the X-ray to \gray\ and the X-ray to optical flux ratios are sensible assuming a neutron star origin for the X-ray flux.
The X-ray source has no cataloged infrared-to-visible counterpart and, through new observations, we set upper limits to its optical emission of $i^\prime >23.0$ mag and $r^\prime >25.2$ mag.
The source exhibits an X-ray spectrum with most likely both a powerlaw and a thermal component.
We also report on the X-ray and visible light properties of the 43 other sources detected in our Chandra  observation.
\end{abstract}


\section{Introduction\label{s:intro}}

Pulsars are often depicted as relatively simple objects: a highly 
magnetized, fast rotating neutron star (NS) emitting radiation along its poles.
Most emission models start from this basic picture, in principle common 
to all isolated pulsars.
The rotation of the magnetic dipole induces an electric field near the 
polar caps in vacuum, which is then short circuited by an e$^\pm$ pair 
plasma (Ruderman \& Sutherland 1975), or in space-charge limited flow, 
which is screened at a pair formation front (Arons \& Scharlemann 1979).
Models seek then suitable regions of particle acceleration.
Although the radio emission is thought to originate in the polar regions 
of the magnetosphere (e.g Michel 1991 and references therein),
the high-energy emission is now thought to 
originate from the outer magnetosphere, with the original polar cap 
concept (Daugherty \& Harding 1996) having been superseded by outer-gap 
(Cheng, Ho \& Ruderman 1986; Romani 1996) and slot-gap (Arons 1983, Muslimov \& Harding 2004) models.
The picture that has emerged is that radiation is emitted in a 
continuous, very broad, spectral range, with curvature radiation 
producing most of the GeV emission (Romani 1996) and synchrotron and/or 
Compton scattering by cascade products producing the near-infrared to 
soft \gray\ emission (Takata \& Chang 2007, Harding et al. 2008).

The high energy emission properties of pulsars were revealed in the 1990s when use of the EGRET experiment on the \textit{Compton Gamma-Ray Observatory} (e.g. Thompson, 2001 and references therein) led to the multiwavelength spectral characterization of half a dozen \gray\ pulsars, including the discovery of the radio-quiet pulsar Geminga (Halpern \& Holt 1992), the second brightest steady GeV source in the \gray\ sky. 
Amongst the EGRET legacy was a sample of 170 unidentified sources, 74 of which are at $|b|<10^{\circ}$ (Hartmann et al. 1999). 
It has now been found that $\approx 43$ EGRET unidentified sources have counterparts in the \Fermi\ Large Area Telescope (LAT) first year catalog (Abdo et al. 2010b; 1FGL) and a large fraction of those lying on the Galactic plane turned out to be pulsars (Abdo et al. 2008, 2009a, 2009b, 2010a, 2010b), a result anticipated by several authors (Yadigaroglu \& Romani 1995, Cheng \& Zhang 1998, Harding \& Muslimov 2005).

In fact, it is very plausible that many more radio-loud or radio-quiet 
pulsars are hidden in the unidentified Galactic LAT sources, although estimates are highly dependent on details of the different emission mechanisms. 
At this point it is still unclear what makes some \gray\ pulsars radio-loud and some radio-quiet. 
Understanding the distinct properties of the individual sources will surely lead to a better understanding of models for the emission mechanisms, for example, the connection between the radio and the near-infrared to the \gray\ component.

The source 2CG 078+2 was one of about twenty \gray\ sources detected by 
the {\em COS-B} satellite 30 years ago (Swanenburg et al. 1981).
The source is in the line of sight towards the supernova remnant G78.2+2.1.
The remnant comprises a 1-degree-diameter, circular, radio shell with 
two bright and broad opposing arcs on its rim (Higgs, Landecker \& Roger 
1977, Wendker, Higgs \& Landecker 1991).
The remnant has a kinetic distance of about 1.5 kpc (Landecker, Roger, 
and Higgs 1980; Green 2009) and is estimated to have an age of 5400 yr 
(Sturner \& Dermer 1995).
2CG 078+2 is often cited as $\gamma$-Cygni due to its proximity to the 
second magnitude foreground star (m$_{\rm V}$ = 2.2, spectral type F8Iab) that lies on 
the eastern edge of the remnant although there is no other physical 
association between the supernova remnant (SNR) and the star.
A small HII region, located close to the star, forms the $\gamma$-Cygni 
nebula.

The \gray\ source, next cataloged as 3EG J2020+4017 (Hartmann et al. 
1999), was suspected to be either extended emission from the 
SNR or a pulsar formed in the supernova event (Sturner \& Dermer 1996).
Soon after its June 2008 launch, the \Fermi\ \gray\ Space Telescope 
highlighted the discovery of a radio-quiet pulsar in the CTA-1 SNR (Abdo et al. 2008) as a first light result.
This was followed by the discovery of twenty-three other \gray\ pulsars 
using blind search techniques (Abdo et al. 2009b, Saz Parkinson et al. 
2010), among them \FL\ lying within the EGRET error circle of 3EG 
J2020+4017. 
\FL\ is a 265-ms pulsar with a spin-down age of 77 kyr and a total spin-down luminosity of 1.1$\times 10^{35}$ erg/s. 
Interestingly, most of the pulsars discovered by the LAT with blind searches have not been seen at other wavelengths.
In fact, only three of the 26 discovered to date have been detected in radio (Camilo et al. 2009; Abdo et~al. 2010c)

Even before the \Fermi\ discovery of pulsations, and because the EGRET 
\gray\ properties of the source matched those of known \gray-pulsars (hard, 
steady and in the Galactic plane), the search for the radio and 
X-ray counterpart of this source began (Brazier et al. 1996; Becker et 
al. 2004, Weisskopf et al. 2006).
No radio pulsar is associated with \FL\ with searches having now been 
conducted down to $100 \mu$Jy at 1665 MHz (Trepl et al. 2010), 
40~$\mu$Jy at 820 MHz (Becker et al. 2004), and to $11 \mu$Jy at 2 GHz (Ray et al. 2011).  
There is an extended radio source (GB6 J2021+4026) in the vicinity which appears to be positioned more or less symmetrically (see Figure 8 of Trepl et al. 2010) about our best position for the X-ray counterpart but there is no evidence that it is associated with the \gray\ pulsar.

We have previously (Becker et al. 2004, Weisskopf et al. 2006) performed 
\Chandra\ observations (ObsIDs 3856 \& 5533) aimed at different portions 
of the $\gamma$-Cygni field and these pointings were based on the best 
known \gray\ positions available at the time.
The latter \Chandra\ observation, ObsID~5533, overlapped with about 3/4 of the {\sl current}
Fermi-LAT 99\%-confidence positional error circle and detected several 
potential X-ray counterparts, including the source designated as "S21" as
reported in Weisskopf et al. (2006).
Subsequently, Trepl et al. (2010) reanalyzed the available \Chandra\ data and also
used XMM-Newton data to search for a counterpart. 
They identify source 2XMM~J202131+402645, 
  at virtually the same location as the \Chandra\ source ``S21'', as the likely counterpart because
the X-ray source position fell within the $\sim$4\arcmin-radius 
 Fermi-LAT 95\%-confidence positional error circle at the time (0FGL, Abdo et al. 2009b). 
 We note that the most recent Fermi-LAT error circle radius (2FGL, Abdo et al. 2011) 
 is 0.6\arcmin\ (\S3) and no longer includes ``S21''. 
We re-evaluate the situation using both the greatly improved LAT source localization 
 and the position measured from timing of \FL.

The current work describes a new observation that completes the \Chandra\ coverage of the field.
As a result of these observations, we conclude that the source originally 
labelled ``S21'' by Weisskopf et al. (2006), 2XMM~J202131+402645 by  Trepl et al. (2010), 
and source \#20 in the new observation, 
remains the most probable X-ray counterpart to \FL.
Moreover, we also show that the X-ray source is dominated by thermal, 
not powerlaw, emission.
In this regard, it is interesting to compare this source with Geminga, 
one of the best studied radio-quiet \gray\ pulsars, which is older (340 
ky) and less luminous ($\sim3\times 10^{34}$ erg/s) than \FL\ but has a similar period and 
period derivative.

\S\ref{s:analysis_image} describes the analysis of the X-ray image;  \S\ref{s:analysis_spectrum} 
describes the analysis of the X-ray spectrum, especially of the X-ray source \#20; \S\ref{s:geminga} compares the X-ray spectral properties of source \#20 to Geminga and CTA 1; \S\ref{s:analysis_timing} describes new Fermi-LAT pulsar timing of \FL; \S\ref{s:optical} describes our search for an optical counterpart; and \S\ref{s:other_sources} discusses properties of the other \Chandra\ sources in the field.
We provide summary conclusions in \S\ref{s:summary}.

\section{\Chandra\ X-Ray Observations and Data Analysis \label{s:xray_obs}}

We obtained a 56-ks \Chandra\ observation (ObsID 11235, 2010 August 27) using the Advanced CCD Imaging Spectrometer (ACIS). 
Here we report the data taken with the back-illuminated CCD ACIS S3 in faint, timed-exposure mode with 3.141-s frame time.
Background levels were nominal throughout the observation. 
Standard \Chandra\ X-ray Center (CXC) processing provided accurate aspect determination.

Starting with level-1 event lists, we reprocessed the data using the CIAO~v4.2 tool {\tt acis\_process\_events} to remove pixel randomization in order to improve the on-axis point spread function (PSF), thus enhancing the source-detection efficiency and positional accuracy.
In searching for sources, we utilized events in pulse-invariant channels 35-550 corresponding 
to 0.5 to 8.0 keV.

To verify the \Chandra\ position accuracy, we compared the pointing parameters (RA, Dec, roll) given by the \Chandra\ data-processing to that using the 19 2MASS sources (Section~\ref{s:oi_2mass}) with high probability X-ray counterparts.
For these calculations we used $0.075\arcsec$ per axis for the 2MASS position 
uncertainty\footnote{http://www.ipac.caltech.edu/2mass/releases/allsky/doc/explsup.html}.
Assuming no change to the \Chandra\ parameters yielded an acceptable positional fit for these 19 
sources with \chisq\ of 24 for 38 degrees of freedom.
Allowing the three pointing parameters to vary marginally improved \chisq\ to 21 for 35 degrees of freedom and would imply the following corrections: $(0.14\pm0.10)\arcsec$ (RA), $(-0.04\pm0.08)\arcsec$ (Dec), and $(43\pm78)\arcsec$ (roll).
However, these corrections, even allowing for errors, being negligible we did not correct the \Chandra\ positions. 

\subsection{X-Ray Image Analysis \label{s:analysis_image}}

\begin{figure}
\epsfig{file=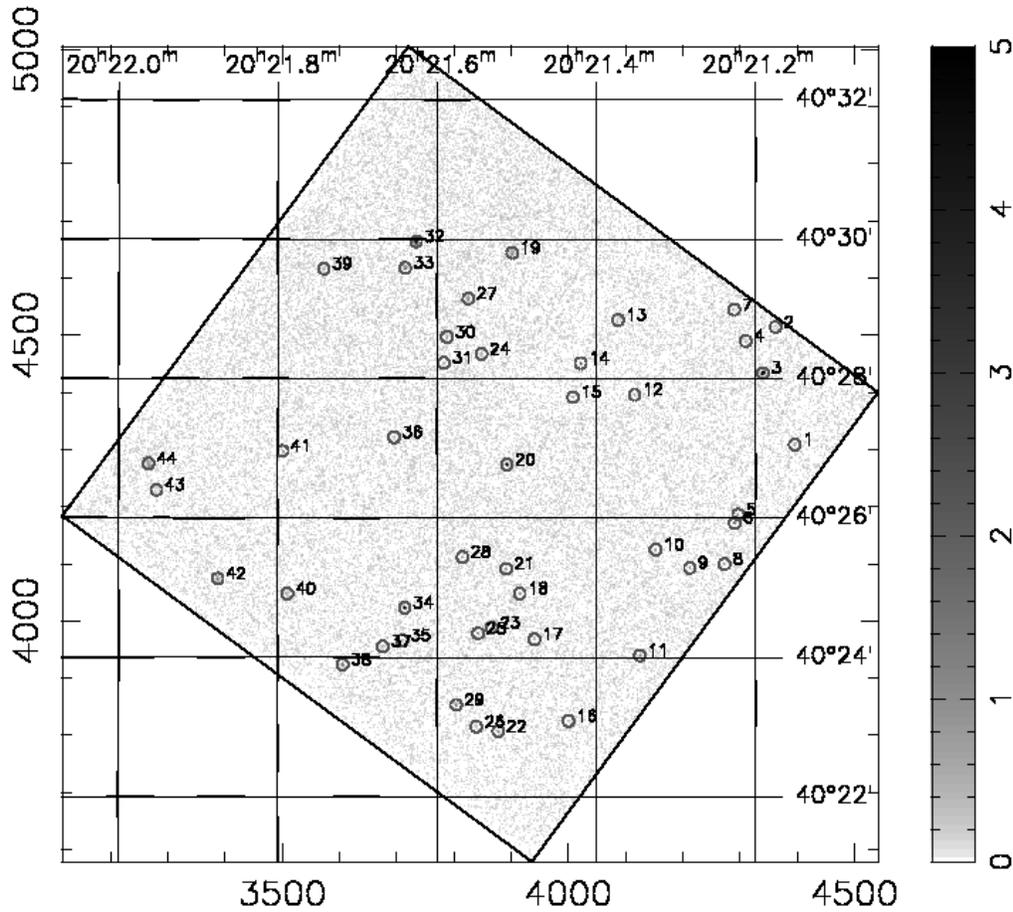,width=12cm,angle=-90}
\figcaption{The field showing the most recent \Chandra\ observation, ObsID~11235. 
For this figure a nearest neighbor smooth has been applied.  
The rotated square shows the extent of the ACIS S3 chip, i.e., the region searched.  
The color bar on the right shows the number of (smoothed) counts detected in
a pixel during the observation.  
The numbers on the left and bottom show the ACIS coordinates in pixels.
Sources are numbered in order of increasing right ascension. See Table~1 for source X-ray properties.
\label{f:new_field}}
\end{figure}

We searched for point-like X-ray sources employing techniques described in Tennant (2006), using a circular-Gaussian approximation to the PSF and setting the signal-to-noise (S/N) threshold for detection to 2.4.
The resulting background-subtracted point-source detection limit is about 7 counts, with fewer than 1 accidental detection expected over the field.
Based upon tests on \Chandra\ deep fields, this approach finds all X-ray sources in those fields down to 10 counts, which we thus regard as the completeness limit.
Figure~\ref{f:new_field} shows the ACIS-S3 image with a small circle at the position of each \Chandra-detected source.

Table~\ref{t:data_x} tabulates the X-ray properties of the 44 {\sl Chandra}-detected sources, with the source number in column 1.
Columns~2--5 give, respectively, right ascension RA, declination Dec, extraction radius $\theta_{\rm ext}$, and approximate number of X-ray counts $C_{x}$ detected from the source.
Column~6 lists the single-axis RMS error $\sigma_{x}=[(\sigma_{\rm PSF}^2/C_{x}) + \sigma^2_{\rm sys}]^{1/2}$ in the X-ray-source position, where $\sigma_{\rm PSF}$ is the dispersion of the circular Gaussian that approximately matches the PSF at the source location and $\sigma_{\rm sys}$ is a systematic error.
Uncertainties in the plate scale\footnote{See
http://cxc.harvard.edu/mta/ASPECT/aca\_plate\_scale/} imply $\sigma_{\rm sys}\approx 0\farcs13$: 
To be conservative, we set $\sigma_{\rm sys}=0\farcs2$ (per axis).
Column~7 gives the radial uncertainty $\epsilon_{99}=3.03\: \sigma_{x}$ in the X-ray position~---~i.e., $\chi^{2}_{2}=9.21=3.03^2$ corresponds to 99\% confidence on 2 degrees of freedom~---~for inclusion of the true source position.
Columns~8 \& 9 are color ratios defined and discussed in \S~\ref{s:analysis_spec}.

In view of the spin-down age and energetics of \FL, the possibility exists that a pulsar wind nebula (PWN)
 may also be present in the X-ray image of the field.
We searched for moderately extended sources in the field and identified three features of interest.
Two are located near the edge of the S3 CCD which increases the likelihood of their being false
positives. The third feature lies $\sim$6\arcsec\ west of source \#20. 
Simultaneously fitting the combination of this feature and source \#20 to circular Gaussians (plus
a constant background) results in a Gaussian width of 2.2\arcsec\ and a total of 22 X-ray events
within this extended feature.

We also compared the spatial distribution of events in and around source \#20 to a model of the 
\Chandra/ACIS PSF valid for its location relative to the aimpoint and its characteristic energy using 
the PSF library psflib v4.1 (CALDB v4.2). Source \#20 is consistent with being pointlike.

\subsection{X-Ray Spectral Analysis \label{s:analysis_spectrum}}

We used the XSPEC (v.12.5) spectral-fitting package (Arnaud, 1996) to perform spectral fits to the  44 X-ray point sources. We treat source \#20 separately in \S \ref{s:s20_spectrum}.
Data were binned to obtain at least 10 counts per spectral bin before background subtraction. 
The background was determined after masking off a circular region around each of the 44 detected X-ray sources corresponding to a circle of radius 25 times the uncertainty listed in column 6 of Table~\ref{t:data_x} and then averaging over the remaining pixels. 
Individual response matrix and ancillary response files appropriate to each source position were
 created for this analysis using the {\tt mkacisrmf} and {\tt mkarf} tools available in CIAO version 4.2.

\subsubsection{The X-Ray Spectrum of Source \#20\label{s:s20_spectrum}}

\begin{figure}
\epsfig{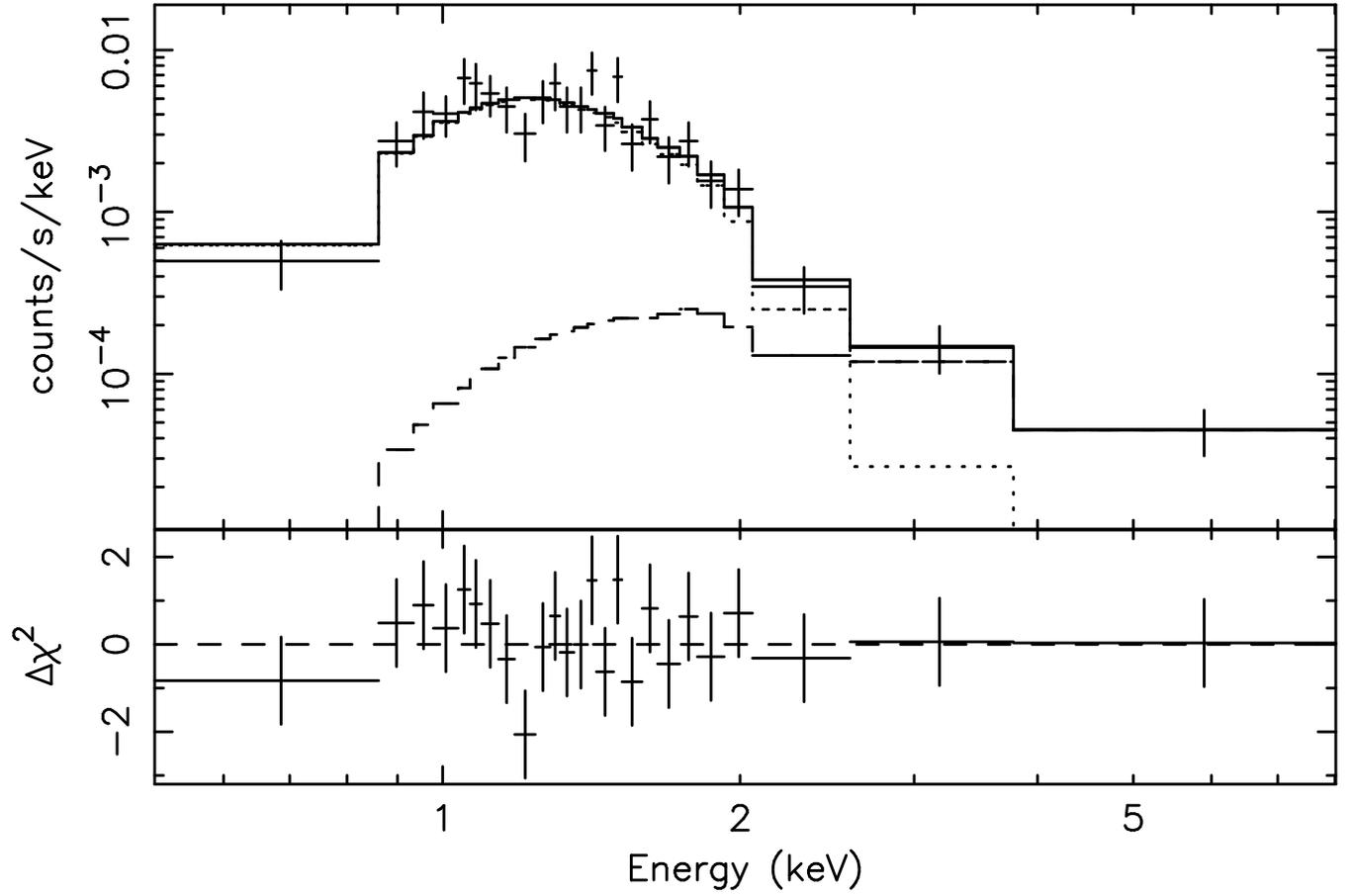}
\figcaption{The spectrum of source \#20 with background subtracted and fit to a {\tt nsa+powerlaw} model.
The dotted line in the upper panel is the {\tt nsa} component, the dashed line is the {\tt powerlaw} component and the solid line is the two components combined.
The lower panel shows the contributions to \chisq. 
\label{f:s20_spectrum}}
\end{figure}

A sufficient number of counts were detected from source \#20 to perform more extensive spectral
 analysis.
Figure~\ref{f:s20_spectrum} shows the X-ray spectrum of source \#20 with background subtracted.
Data were again binned to obtain at least 10 counts per bin before background subtraction. 
The background is less than 3\% of the signal plus background from the region that includes source \#20. 

We begin our spectral analysis by first considering single-component spectral models (Table~\ref{t:smodels}) with a multiplicative absorption component: an absorbed power law ({\tt powerlaw} in XSPEC); an absorbed black body ({\tt bbodyrad}); and three different absorbed neutron star atmosphere models,\footnote{A NS radius of 12.996~km and mass of 1.358~M$_{\odot}$ is assumed throughout for purposes of computing effects of gravitational redshift in the neutron star atmosphere models} namely, {\tt nsa} (Pavlov et al. 1995), 
{\tt nsmax-1260} and {\tt nsmax-130190} (Ho, Potekhin \& Chabrier 2008).

For computing absorption, we utilized abundances (XSPEC's {\tt wilm}) from Wilms, Allen, \& McCray (2000) with cross-sections ({\tt vern}) from Verner et al.(1993) and allowed for interstellar extinction by grains using the model ({\tt tbabs}) of Wilms, Allen, \& McCray (2000).

All of these models provide statistically adequate fits to the data in the 0.5-8.0 keV range (Table~\ref{t:smodels}).
(We note that so did a fit to an absorbed {\tt mekal} model.)
From the {\sc Hi} in the Galaxy (Dickey \& Lockman, 1990), one infers a column density $N_{\rm H} \approx 1.4\times 10^{22}$~cm$^{-2}$ through the Galaxy in the direction of source \#20, implying that values below this should be expected if there is no circumstellar absorption, and this is indeed the case (Table~\ref{t:smodels}). 
Next we posit that the very steep power law index of almost 5 is not physical, but indicative of a soft, thermal component. 
The {\tt bbodyrad} model's normalization,
$(R^2_{\rm km}/D^2_{10{\rm kpc}})=0.80$, where $R_{\rm km}$ is the radius of the emitting area in km and $D_{10{\rm kpc}}$ is the distance to the source in units of 10~kpc,
corresponds to an emitting area of $R_{\rm km}=0.13$, assuming the association of the pulsar with G78.2+2.1 at a nominal distance of $D_{10{\rm kpc}}=0.15$. 
This is much smaller than a typical NS radius, but not very different than the standard polar cap radius,
$\approx R_{NS} sin \theta \approx R_{NS}^{3/2} (2\pi/cP)^{1/2}$ which for $P=0.265$~s and $R_{NS}=13$~km is 0.42~km.

Conversely, if we assume the emission comes from a NS with a characteristic radius of 13 km, then the {\tt bbodyrad} norm implies a source distance of $\approx145$ kpc, well outside the Galaxy. 
This distance estimate drops for the different NS-atmosphere models, e.g. down to 13 kpc (Table~\ref{t:smodels}), still somewhat distant to remain within the Galaxy.
In addition, the temperature estimates of the (cooling) NS, range from $\log(T_{\infty})\geq 6.5$ to $6.1$, with the precise value depending on the model (Table~\ref{t:smodels}).
These estimates are consistent with those expected for a pulsar of an estimated age somewhere between 5400 and 77000 years, depending of course on the equation of state, composition of the heat-blanketing envelope, and the degree of superfluidity in the star's core.
(See Yakovlev et al. 2010 and references therein for recent details on the subject of cooling NSs.)

We then ask whether or not combining a power law with the other models is indicated by the data. 
Table~\ref{t:dmodels} tabulates the change in \chisq, the f-statistic, and the probability that combining models has significantly improved the quality of the fit. 
The table also tabulates he derived spectral parameters.
In all cases combining various thermal models with a power law does improve the quality of the fit with a confidence better than 2-sigma, but not 3-sigma.
Moreover, these two-component models allow a wide latitude for the uncertainties of the best-fit parameters and hence these parameters are not as constrained as one might wish.
This follows from the fact that all of the single models (Table ~\ref{t:smodels}) themselves provide statistically adequate fits. 
Thus, the 3-sigma contours for any two-component model allows for one or the other of the component models to have a zero norm -- e.g. a power law component is not completely {\em required} by the data at this level of significance.
Keeping this proviso in mind, we continue to examine the two-component models. 
Certainly the physical interpretation of the data is perhaps more sensible when both components are introduced, especially in light of the strong \gray\ emission which cannot arise from any single-component ``thermal'' model spectrum.
In addition, both the inferred NS temperatures and distances implied by the two-component models (Table~\ref{t:dmodels}), with the possible exception of the {\tt bbodyrad+powerlaw} model,
are consistent with what one might expect for a young cooling neutron star within the galaxy.

\subsubsection{Discussion of the Spectrum of Source \#20 \label{s:spectrum_discussion}}

The considerations of the single-component models in the previous section lead one to conclude that we are perhaps seeing a hot spot of size comparable to the polar cap rather than thermal emission from the entire surface of the neutron star and with a temperature higher than one expects from cooling of a neutron star at an age $> 5000$ years.
It is thus possible that the bulk of the emission comes from heating of the polar cap by backflowing accelerated particles.
The expected luminosity and temperature from heating by positrons produced by curvature radiation of primaries in a space-charge limited flow model is $L_+ \sim 10^{31}\,\rm erg\,s^{-1}$ and $\log(T_+) \simeq 6.3$ (Harding \& Muslimov 2001).
This luminosity is deposited over an area roughly that of the polar cap and radiated on a timescale less then the heat diffusion timescale across field lines to other areas of the NS.  
The temperature $T_+$ is close to that determined from the {\tt bbodyrad} and {\tt nsa} model fits allowing for the gravitational redshift.
It is therefore quite possible that the X-ray emission from PSR J2021+4026 is dominated by polar cap heating.

On the other hand, 
many well-studied neutron stars that exhibit both X-ray and \gray\ emission have composite X-ray spectra, showing non-thermal power law magnetospheric emission and/or hot polar cap emission in addition to the lower-temperature full surface thermal emission (e.g. Geminga, PSR B0665+14, PSR B1055-52, see De Luca et al., 2005).
While the present data quality do not demand such two-component models, the inferred distances from such two-component fits, e.g., $\sim 6$~kpc for the NS atmosphere models (Table ~\ref{t:dmodels}) become reasonable for plausible stellar radii assuming full surface emission. This distance is larger than
 the kinematic distance to the SNR~78.2$+$2.1 of 1.5~kpc but is comparable to the distance to the 
 Cygnus arm at Galactic longitude $\sim$78\arcdeg.
Thus, it is possible that a two-component model, with full-surface cooling and magnetospheric power law emission present at lower levels, might be needed to accurately describe the emission
physics.
In practice, decomposing such complex X-ray spectra requires good statistics and phase-resolved X-ray spectroscopy. 
For a target this faint, extremely long observations or next-generation X-ray satellites are clearly required.

\subsection{Comparison of the Spectrum of \#20 to Geminga and CTA 1 \label{s:geminga}}

There are similarities and differences between the X-ray spectrum of \FL\ and two of the other radio-quiet \gray\ pulsars with detected X-ray emission, Geminga and CTA 1. 
First \FL, like Geminga (Jackson and Halpern, 2005) and CTA 1 (Caraveo et al. 2010), may also be characterized by two spectral components, a thermal component and power law.
In Geminga, however, the power law component begins to dominate above about 0.5 keV and $\log (T_{\infty})$ is about 5.7. 
For \FL\ the power law dominates above 2.5 keV and $\log (T_{\infty})$ is higher, as one would expect as \FL, based on its spin-down age, is younger. 

The spectral indices for the power law components for source \#20 and Geminga are not dissimilar, but one needs to recognize the large uncertainty in the measurements reported here.
Another, possibly important, spectral difference between the two X-ray spectra is that, in the case of Geminga, the blackbody component gives an emission radius that is plausible for a NS radius.
This is not so for \FL. 
In this case, the neutron star atmosphere models seem to yield more physically reasonable parameters than the {\tt bbodyrad+powerlaw} model.
If we assume that the younger and hotter star still has an atmosphere while the older Geminga does not, then these results are sensible. 
Finally, there is a weak extended emission feature near source \#20 that may be indicative of a PWN. If so, then it extends no more than 0.04 to 0.17~pc from source \#20 (assuming a distance of 1.5 to 6.0~kpc, respectively) and contributes $\sim$7\% of the X-ray counts  detected from source \#20 and its surroundings. 
Emission associated with the Geminga PWN has a similar extent and contributes ~10\% of the non-thermal X-ray flux of the pulsar, but only about 1\% of the total flux (Pavlov \etal\ 2010).

For CTA 1, the measured temperature, powerlaw index and emission radius are $\log (T_{\infty}) = 6.08$, $\Gamma = 1.3$ and $r = 0.64$ km for a {\tt powerlaw+bbodyrad} model, and $\log (T_{\infty})= 5.78$, $\Gamma = 1.25$ and $r = 4.92$ km for {\tt powerlaw+nsa}, with slightly lower $\chi^2$ for the former (Caraveo et al. 2010). 
In both cases, the emitting radius is significantly smaller than a standard NS radius.  
Thus, similar to \FL, CTA-1 shows a possible heated polar cap component, with a temperature very close to the model prediction of $\log (T_{\infty}) \simeq 6.2$ (Harding \& Muslimov 2001), however, the emitting radius in this case is a factor of 2.5 larger than the polar cap radius.
CTA 1 does not show evidence for a cool component, with the upper limits making it unusually cool for its age.

\section{\Fermi-LAT Localization and Timing Analysis\label{s:analysis_timing}}

The \Fermi-LAT normally localizes \gray\ sources using the incident \gray\ photon directions. 
The LAT has a point spread function that is strongly energy dependent, with a resolution of about 0.8$^\circ$ at 1 GeV. 
For a bright source, however, localization with arcminute accuracy is possible. 
The source in the second LAT catalog (Abdo et al. 2011) that corresponds to \FL\ is 2FGL J2021.5+4026.  
The catalog position for this source is R.A. 20$^h$21$^m$34$^s$.1, Decl.\ $+$40\arcdeg26\arcmin28\arcsec\ (J2000), with a 95\% confidence radius of 0.60\arcmin.

For pulsars, one can use timing techniques better to localize the source, independent of the photon direction measurements, as described in Ray et al. (2011).  
The position determination from timing of this pulsar is hampered by the large contribution from rotational instabilities common in young pulsars and manifest as ``timing noise''.
Therefore, we have taken two different approaches to try to confirm the association between \FL\ and source \#20. 
For this analysis, we used \grays\ detected by the LAT from 2008 Aug 4 to 2011 Mar 12, selecting only those within $0.8^\degree$ from the previous best position (Ray et al. 2011), and with energies greater than 400 MeV. 
These ``cuts'' were chosen to maximize the significance of the pulsation. 
We chose only photons belonging to the most restrictive ``diffuse'' class according to the ``Pass~6'' instrument response functions (see Atwood et al. 2009), which have the lowest background contamination. Furthermore, we selected only photons with a zenith angle of $<105^\degree$, to reduce contamination due to  secondary-atmospheric \grays.

\begin{figure}
\epsfig{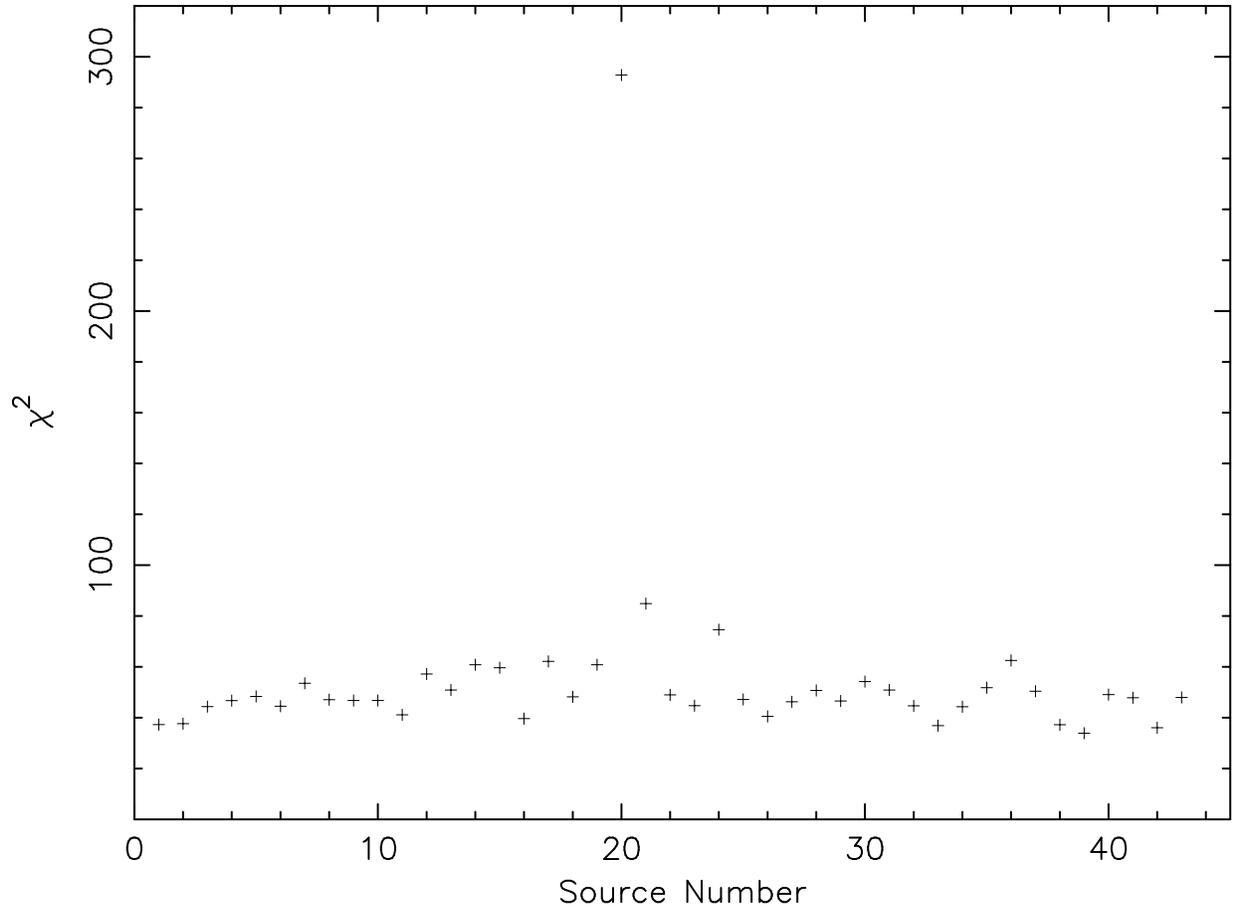}
\figcaption{Significance of pulsation detection, as measured by the \chisq\ test, versus source number.
The data were binned into 20 pulse phase bins and fit to a constant flux model (no pulsations). 
The worse the fit (higher \chisq), the more likely the X-ray source is the \gray\ pulsar counterpart.
\label{f:pulse_timing}}
\end{figure}

As a first test, we evaluate the significance of the \gray\ pulsations by assuming the pulsar is at each of the candidate X-ray source locations seen in our Chandra observation (Table ~\ref{t:data_x}). 
For each candidate location, we transform arrival times to the barycenter using the X-ray position and then use the {\tt prepfold} routine from the PRESTO pulsar package (Ransom, Eikenberry \& Middleditch 2002) to find the frequency ($f$), and its first and second derivatives, $\dot{f}$ and $\ddot{f}$, which maximize the statistical significance of the pulsation.
Figure~\ref{f:pulse_timing} shows the results from this exploratory search, where it is clear that source \#20 gives the maximum significance for pulsation using this algorithm.  
This indicates that of the possible X-ray sources in the field, source \#20 is the most likely X-ray counterpart.

\begin{figure}
\epsfig{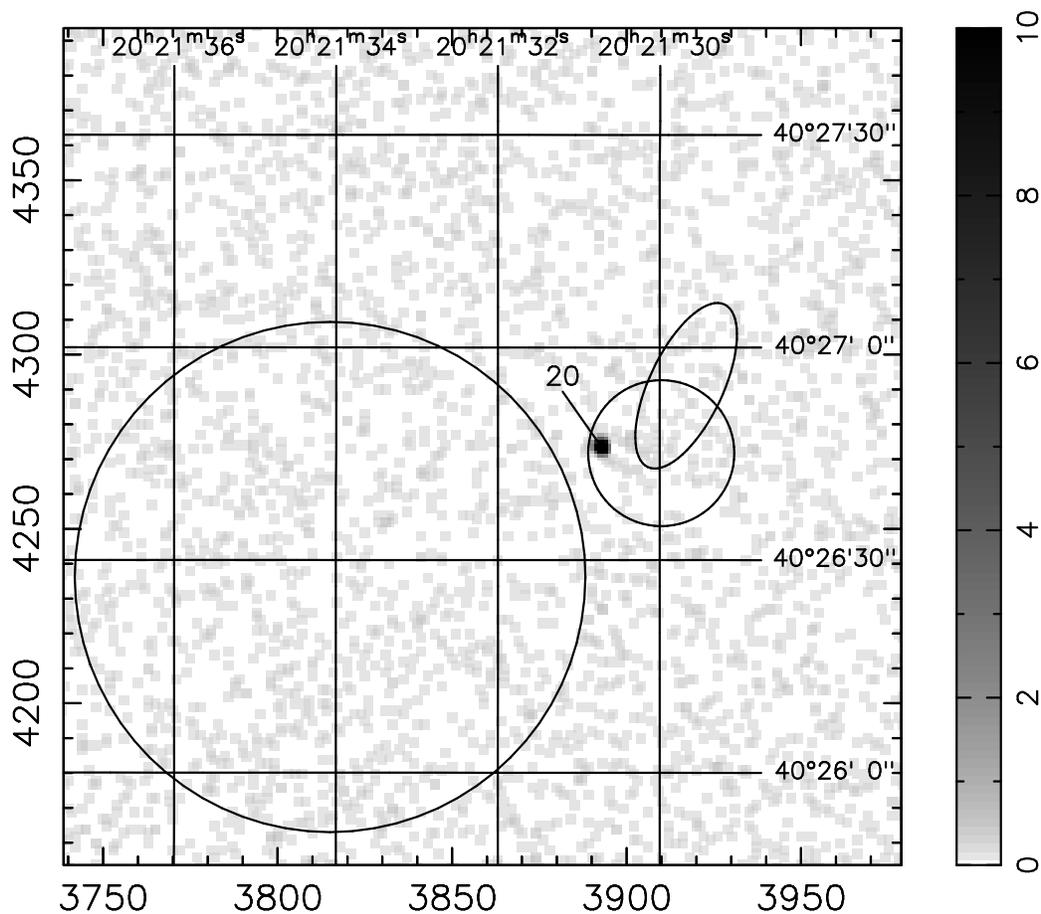}
\figcaption{Same as Fig.~\ref{f:new_field} but now just showing the region around
source \#20.  
The large circle denotes the most recent LAT imaging position 95\% confidence error circle (2FGL, Abdo et al. 2011). 
The small ellipse is the 95\% confidence timing ellipse from the current
work using WAVE terms to estimate the impact of timing noise.
The small circle is the 95\% confidence region obtained by combining the 
timing solution determined by Ray et~al. (2011) added in quadrature with our new (10\arcsec) estimate of the systematic error. See text for further details.
\label{f:snapshot}}
\end{figure}

Next, we use pulsar timing to fit for the position of the pulsar, as described in Ray et al. (2011).
We measured 55 times of arrival (TOAs) based on 22-day integrations spanning the data set described above. 
The typical uncertainty on each TOA measurement was 4.7 ms.
Using \textsc{Tempo2} (Hobbs et al. 2006), we fit the TOAs to a timing model including $f$, $\dot{f}$, and $\ddot{f}$. 
With only these terms in the model, we observe very large residuals and the $ \chi^2 $ of the fit is very poor. 
This poor model fit means that the statistical errors in the fitted right ascension and declination reported by \textsc{Tempo2} are unreliable. 
To get an estimate of the systematic error in the position fit, we use the following procedure.
We added 5 so-called ``WAVE'' terms to the timing model to account for the timing noise using harmonically-related sinusoids (Hobbs et al. 2004). 
We then perform a fine scan over a positional grid around the location of source \#20.
Holding the position fixed at each grid point, we fit for the spin and WAVE parameters. 
The grid position with the lowest resulting $\chi^2$ for the fit is R.A. 20$^h$21$^m$29$^s$.683,  Decl.\ $+$40\arcdeg26\arcmin54.61\arcsec\ (J2000).  
This new timing position is $10\arcsec$ away from the one reported in Ray et al. (2011) which was based on 14 fewer months of data. 
Based on the $\chi^2$ map over the grid, we estimate the 95\% confidence region 
of the new timing position to be an ellipse of dimensions 26\arcsec$\times$10\arcsec, as shown in Figure~\ref{f:snapshot}.
The separation between the position of source \#20 and the refined timing analysis position 
 obtained here is $14.7\arcsec$ and source \#20 lies outside the 95\% confidence region.

\begin{figure}
\epsfig{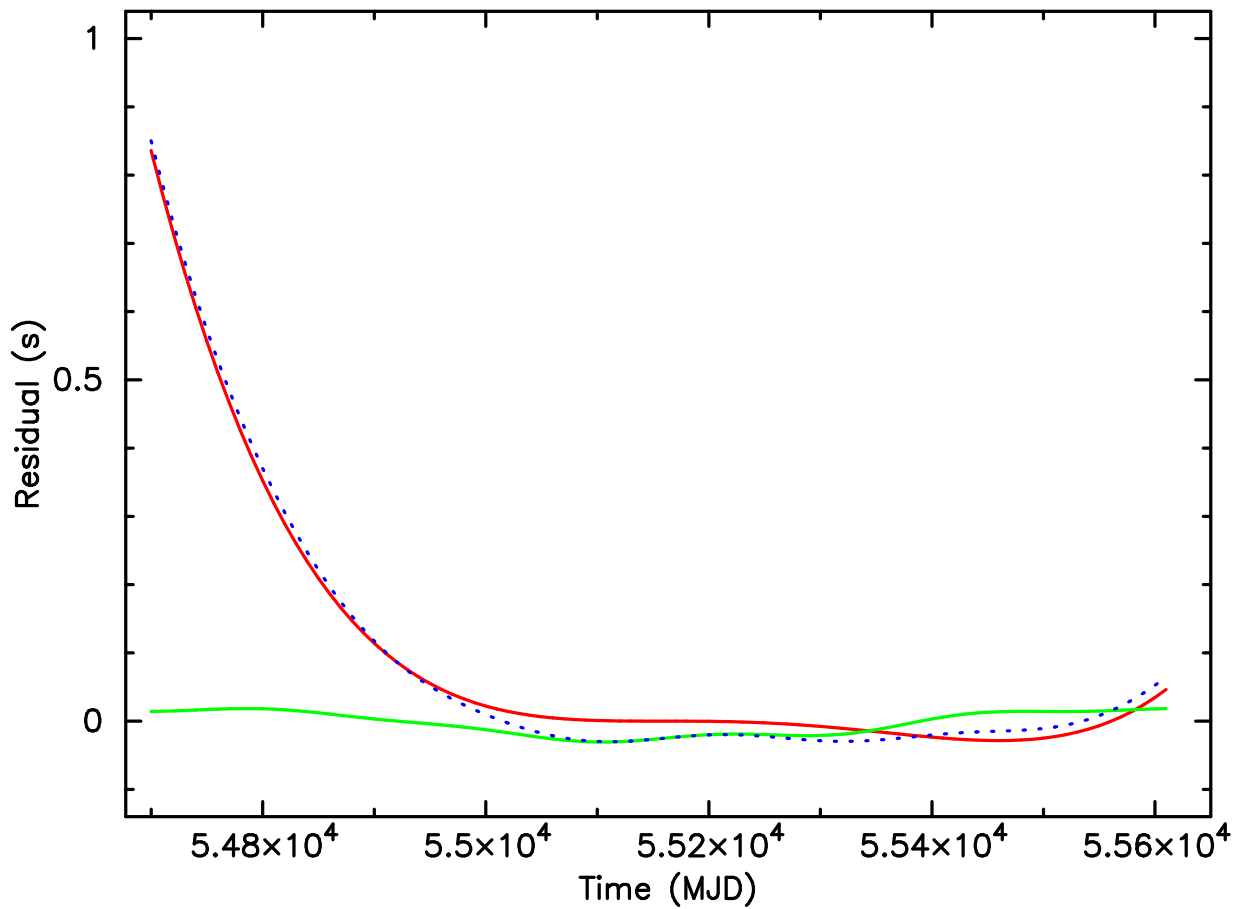}
\figcaption{Illustration of the magnitude of the timing noise observed in \FL. The dotted (blue) curve shows the timing residuals that can be attributed solely to timing noise.  
This is the sum of two components: The thin upper (red) curve is the contribution from the polynomial terms $\ddot{f}$ and $\dddot{f}$, neither of which can be attributed to the secular spin down of the pulsar. The lower (green) curve is the combined sinusoidal WAVE components (see text). Note that the pulsar period is 0.265 seconds, so timing noise causes several complete phase wraps over this interval.
\label{f:tnoise}}
\end{figure}

A precise evaluation of systematic timing errors is complicated by the (erratic) timing behavior of the pulsar itself and the relatively short data span with respect to the 1-year modulation that is introduced by an incorrect position. 
That is, the position can be perturbed by any component of the timing noise that appears to be a 1-year sinusoid. 
The magnitude of this effect is difficult to estimate because we have just one realization of the stochastic timing noise process in our data.  
To see the potential contribution, we have plotted an estimate of the timing noise contributions in Figure~\ref{f:tnoise}.  
Here, we assume that the parameters $f$ and $\dot{f}$ are dominated by the secular spindown of the pulsar, while any higher order frequency derivatives and all of the WAVE parameters are dominated by timing noise. 
Clearly, this is a red-noise stochastic process. 
Note that a position error of 10$\arcsec$  will introduce a sinusoidal term with a 1-year period and an amplitude of 24 ms.
To get another estimate of the systematic error, we fit a timing model including the pulsar position to three overlapping segments of data, the first half of our data span, the middle half, and the last half. These three fitted positions are separated by 7--9$\arcsec$, giving another estimate of the systematic error resulting from timing noise. 

Based on these considerations, we adopt $10\arcsec$ as an estimate of the systematic error in the timing position. Combining, in quadrature, this systematic error estimate with the error estimated in Ray et~al. (2011) results in the smaller, 10.3\arcsec\ radius, circular 95\% confidence region depicted in Figure~\ref{f:snapshot}. Source \#20 lies within this region.

In summary, when using pulsar timing to derive a position at the few arcsec level it is important to allow for low frequency (year timescale) timing noise.
We have done this two ways: First, by adding WAVE terms to the solution and allowing these terms to vary when deriving an error in our grid search.
Second, by time slicing the data and looking at how the derived position changes. 
As seen in Figure~\ref{f:snapshot}, each of these methods gave similar sized error regions that overlap.

\section{Search for an Optical Counterpart to the X-Ray Source\label{s:optical}}

As reported in Weisskopf et al (2006) and repeated in \S\ref{s:oi}, there are no cataloged optical counterparts for source \#20. 
This is not surprising as, with the exception of the $m\approx 16$ Crab, most optically detected pulsars have magnitudes $\geq 25$.
In addition the field of \FL\ is crowded and optical observations are further hampered by the presence of the 8$^{th}$ magnitude star BD+39 4152 (=V405 Cyg) one arcmin away to the east. 

We present here observations of the field taken on October 31, 2008 with the OPTIC orthogonal frame-transfer camera on the Kit Peak National Observatory, 3.6m, Wisconsin, Yale, Indiana, \& NOAO (hence WIYN) telescope. 
The OPTIC camera, with a 10$^\prime$ field and plate scale 0.141$^{\prime\prime}$/pixel, allows improved image quality 
through ``Orthogonal Transfer'' (OT) rapid electronic guiding following motions of a reference star (Tonry et al. 2004).
We used BD+39 4152 itself as the guide star and were able to correct at 50\,Hz, collecting $3\times180$\,s dithered exposures in $r^\prime$ and $i^\prime$.
These frames were subject to standard calibrations, except for the flat field
frames which were assembled by applying image shifts matching those of the
OT guiding during the individual science exposures.
The resulting image stacks have final PSF widths of 0.87$^{\prime\prime}$ ($r^\prime$) and 0.62$^{\prime\prime}$ ($i^\prime$) near the guide star; the PSF width increases by $\sim 30$\% toward the edge of the frame. 
We estimate that the frame is aligned to the \Chandra\ coordinates with $<0.2^{\prime\prime}$ precision.

\begin{figure}
\begin{center}
\epsfig{file=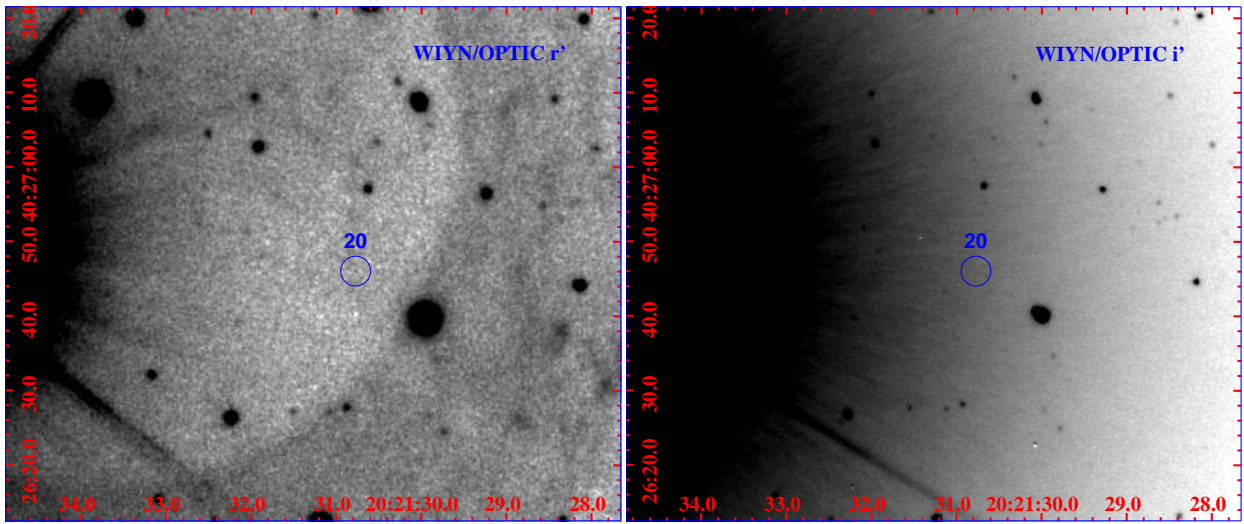,width=16.5cm,angle=0}
\end{center}
\figcaption{\label{f:WIYN_r2_i2} 
Portions of the WIYN 3.6m/OPTIC $r^\prime$ (left) and  $i^\prime$ (right) images of the field of \FL.
Reflected light and scattering from the 8$^{th}$ magnitude star BD+39~4152, used here as a guide star and located to the left of each image, dominates the fields. 
The position of \Chandra\ source \#20 is marked with a circle of radius 2 arcsec. 
There is no detection visible in either band with upper limits as given in the text.
}
\end{figure}

Figure~\ref{f:WIYN_r2_i2} shows a portion of the OPTIC frames in $r^\prime$ and $i^\prime$ centered near the position of source \#20. 
Note the secondary reflections of BD+39 4152 and the strong scattering background, especially in $i^\prime$.
Magnitudes were corrected to the SDSS photometry scale using observations of the calibration star Ru 149F (Smith et al. 2002).
We measured the fluxes of the faintest detectable stellar sources in the
vicinity of our target position and used these to estimate upper limits 
($\sim 95$\% confidence) on the undetected optical flux for source \#20 
of $i^\prime > 23.0$ mag (the sensitivity is severely limited by scattered flux) and $r^\prime > 25.2$~mag.

Some of the diffuse emission toward the right (west) of the $r^\prime$ image is part of larger scale filamentary structure visible over several arcminutes. 
This is likely H$\alpha$/[NII] associated with the \gcyg\ SNR itself.
We note that this remnant has been poorly studied in the optical.  
Mavromatakis (2003) described extensive diffuse line emission over 
a $\sim 1^\circ$ region, but found little filamentary emission and was not able to detect the very faint filaments seen in our data.  
No corresponding X-ray structure is seen, supporting the claim in Mavromatakis (2003) that the \gcyg\ remnant is dominated by low velocity shocks.  
Deep narrow band imaging to trace this structure could be useful in testing the connection, if any, between PSR J2021+4026 and the \gcyg\ SNR.

\section{The Other 43 X-Ray Sources in the Field \label{s:other_sources}}

\subsection{Spectral Analysis\label{s:analysis_spec}}

\begin{figure}
\begin{center}
\epsfig{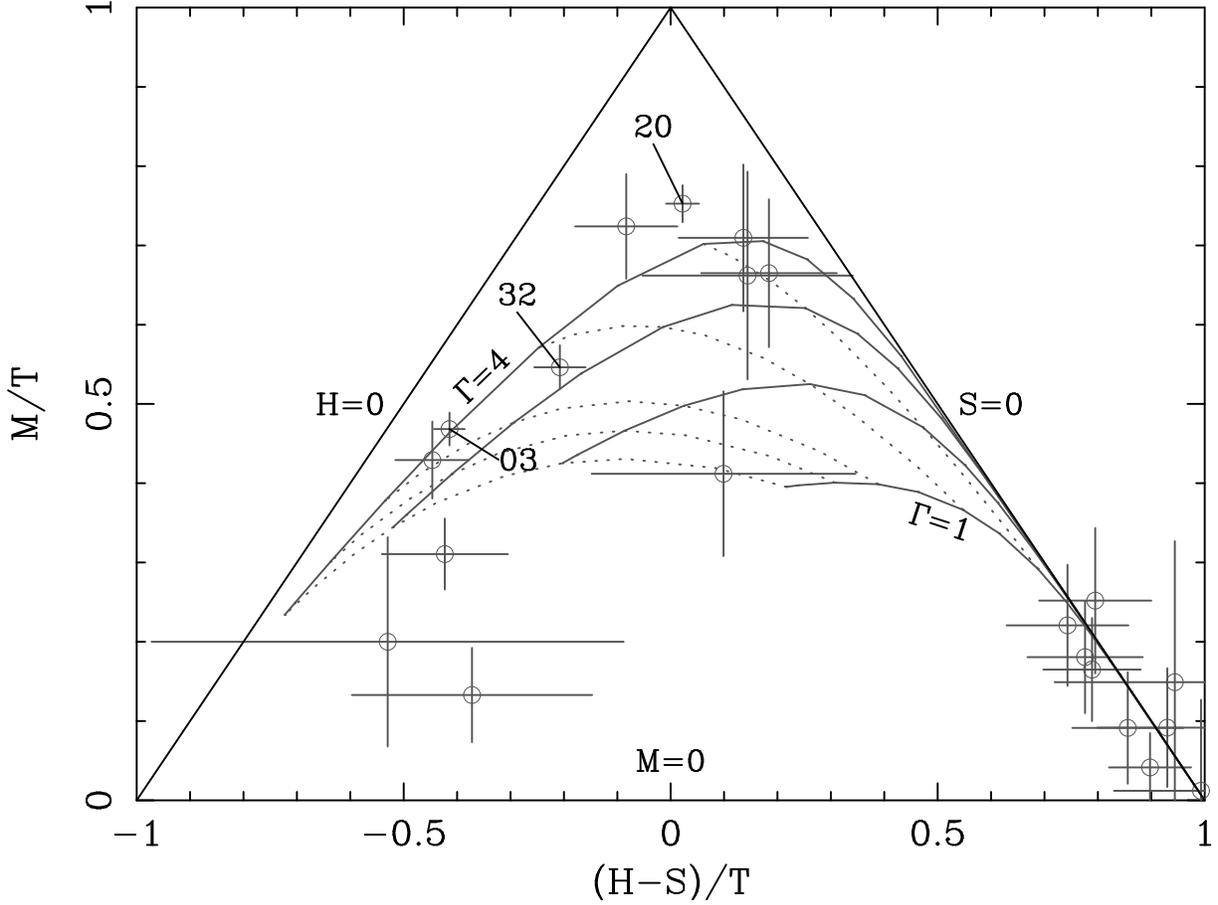}
\end{center}
\figcaption{The X-ray color-color diagram for the 21 X-ray sources that have more than 15 source counts.  
The Chandra band is divided into 3 bands with S (soft) covering 0.5 to 1.0 keV, M (medium) covering 1.0 to 2.0 and H (hard) 2 to 8 keV.  
The Total, T, is the sum of S, M and H.  
The solid curves within the triangle 
represent powerlaw spectra with photon indices of 1, 2, 3 and 4.  
The dotted curves correspond to lines of constant absorbing columns of 
({\sl bottom to top}) $1\times 10^{20}$,
$1\times 10^{21}$, $2\times 10^{21}$,$ 5\times 10^{21}$, and $1\times10^{22}$, respectively. 
The triangle encloses the physically-meaningful range of colors. 
See text for the discussion
\label{f:hratio}}
\end{figure}

There is a drop in the \Chandra\ energy response above the mirror coating's iridium-M edges ($\approx\!2$~keV), thus any source with a substantial fraction of its detected photons above 2~keV is indicative of a very hard spectrum.
Figure~\ref{f:hratio} shows the X-ray color-color diagram for the 21 sources that have more than 15 source counts.  
The diagram comprises 3 bands: S (soft) covering 0.5 to 1.0 keV; M (medium) covering 1.0 to 2.0 and H (hard) 2 to 8 keV with T (total) simply the sum of S, M and H.
The color ratios that comprise Figure~\ref{f:hratio} are given in columns 8 \& 9 of Table ~\ref{t:data_x}.
The X coordinate in Figure~\ref{f:hratio} (H-S)/T measures how hard the spectrum is and the y coordinate (M/T) measures how centrally peaked the spectrum is. 
Positive source counts requires data points to be inside the triangle, but background subtraction causes a few to appear slightly outside.
Sources with (H-S)/T greater than 0.5 are spectrally very hard and are likely background AGNs shining through the galactic plane.  
Only one of this group of sources, \#42, has a cataloged optical counterpart (Table~\ref{t:data_oi}).
Sources with negative values of (H-S)/T likely have thermal spectra and are plausibly lightly absorbed foreground stars.
Note that source \#20 has the highest fraction of counts in the 1-2 keV
band. 
Two X-ray sources, other than source \#20, have sufficient counts to warrant a spectral analysis: \#3, the brightest in the field; and \#32.

\subsubsection{The X-Ray Spectrum of Source \#3\label{s:s3_spectrum}}
We first fit the source~\#3 data using absorbed {\tt powerlaw}, {\tt bbodyrad}, and {\tt mekal} models. 
None of these models provided acceptable fits to the data, \chisq\ being 48.3, 57.2, 70.8, respectively, for 32 degrees of freedom. 
The only two-component+ model that provided an acceptable fit was a two-temperature {\tt mekal} model with \chisq\ of 23.4 for 30 degrees of freedom, further indicating that this is a foreground star. 
The results of our spectral analysis, following the procedures discussed in \S\ref{s:s20_spectrum}, are in Table~\ref{t:s3_spectrum}.

\subsubsection{The X-Ray Spectrum of Source \#32\label{s:s32_spectrum}}
The data for source \#32 are well-fit by an absorbed powerlaw model (\chisq\ of 15.7 for 19 degrees of freedom), but, as with source \#20, the powerlaw index is very steep being 3.4. In this case, however, neither a single-temperature mekal model nor a {\tt bbodyrad} model provide as compelling fits (\chisq\ of 46.9 \& 23.9 respectively), although the latter is statistically acceptable. 
There is simply too much uncertainty to firmly classify this source on the basis of the X-ray spectroscopy alone. 

\subsection{Temporal Variability\label{s:analysis_var}}

The general paucity of counts also precludes a sensitive time-variability analysis for almost all these sources.
Nonetheless, one of the three X-ray-brightest sources, \#32, shows evidence for a significant temporal variation, suggestive of stellar coronal emission.
The existence of both a likely 2MASS and USNO candidate counterpart (Table~\ref{t:data_oi}) reinforces this interpretation in which case simple spectral models might not be expected to fit the data, as we have seen (\S\ref{s:s32_spectrum})

\subsection{Candidate Catalog Optical and Near-Infrared Counterparts\label{s:oi}}

We searched for candidate optical counterparts to the detected X-ray sources.
We used HEASARC's {\tt BROWSE}\footnote{See
http://heasarc.gsfc.nasa.gov/db-perl/W3Browse/w3browse.pl.} 
feature to search for cataloged objects within the 99\%-confidence radius ($\epsilon_{99}$) of X-ray source positions in Table~\ref{t:data_x}.
Table~\ref{t:data_oi} tabulates results of a cross correlation of the X-ray positions of the \Chandra-detected sources (column 1) 
with optical sources (columns~2--7) in the USNO-B1.0 catalog (Monet et al. 2003) and with near-infrared sources (columns~8--12) in the 2MASS catalog (Skrutskie et al. 2006).

\subsubsection{USNO-B1.0\label{s:oi_usno}}

For fourteen (14) X-ray sources, we found a USNO-B1 (optical) source within the 99\%-confidence radius $\epsilon_{99}$ of the \Chandra\ position (Table~\ref{t:data_x}).
Table~\ref{t:data_oi} columns~2--4 list, respectively, the USNO-B1 right ascension RA, declination Dec, and RMS positional error $\sigma_{o}$ in the form ($\sigma_{o}({\rm RA})$, $\sigma_{o}({\rm Dec})$).
Column~5 gives the angular separation $\delta_{ox}$ between optical and X-ray positions; column~6, the I-band magnitude.
Column~7 estimates the probability $p_o(\delta_{ox}, {\rm I})$ for a chance coincidence within the observed separation of an object as bright or brighter than the I magnitude of the optical candidate.
We determined this probability from the I-magnitude distribution of the 893 USNO sources within $6\arcmin$ (slightly larger than the 8$\times$8\arcmin\ \Chandra\ field of view)
 of the X-ray pointing direction.
We designate a potential optical counterpart to an X-ray source as a `strong candidate' only if the sample impurity---i.e., probability of chance coincidence---$p_o(\delta_{ox}, {\rm I})<1\%$.
All the candidate USNO-B1 sources satisfy this criterion.

\subsubsection{2MASS\label{s:oi_2mass}}

For nineteen (19) X-ray sources, we found a 2MASS (near-infrared) source within the 99\%-confidence radius $\epsilon_{99}$ of the \Chandra\ position (Table~\ref{t:data_x}).
Table~\ref{t:data_oi} columns~8 and 9 list, respectively, the 2MASS right ascension RA and declination Dec, each with an RMS positional error $\sigma_{i}\approx 0\farcs080$.
Column~10 gives the angular separation $\delta_{ix}$ between near-infrared and X-ray positions; column~11, the K$_{s}$-band magnitude.
Column~12 estimates the probability $p_i(\delta_{ix}, {\rm K}_{s})$ for a chance coincidence within the observed separation of an object as bright or brighter than the K$_{s}$ magnitude of the infrared candidate.
We determined this probability from the $K_s$-magnitude distribution of the 1188 2MASS sources within $6\arcmin$ of the X-ray pointing direction.
We designate a potential near-infrared counterpart to an X-ray source as a `strong candidate' only if the sample impurity---i.e., probability of chance coincidence---$p_i(\delta_{ix}, {\rm K}_{s})<1\%$.
Eighteen (18) sources satisfy this criterion, the exception being source \#44.
Note that the 2MASS set of 18 strong candidate counterparts includes 13 of the 14 USNO-B1 set of strong candidates (\S\ref{s:oi_usno}).

\begin{figure}
\begin{center}
\epsfig{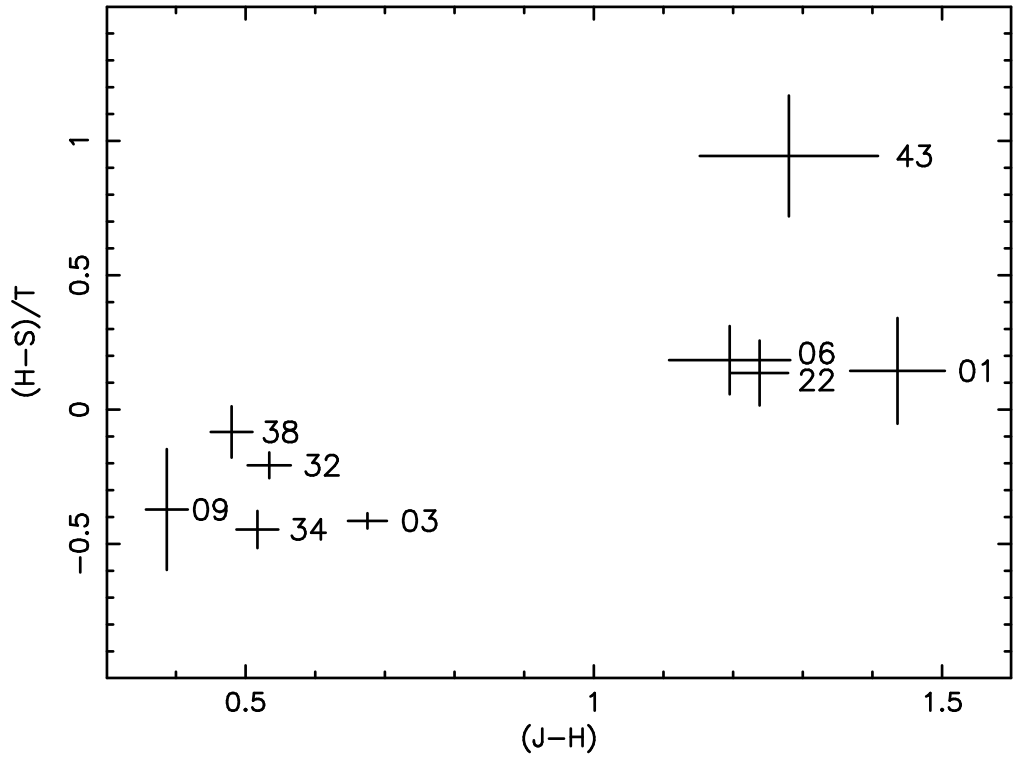}
\end{center}
\figcaption{X-Ray hardness ratio (H-S)/T defined in the caption to figure ~\ref{f:hratio} versus the near infrared color J-H for X-ray sources with 15 or more counts and likely 2MASS candidate counterparts.
\label{f:2MASS_colors}}
\end{figure}

Table~\ref{t:2MASS_colors} tabulates the 2MASS near-infrared photometry (columns~2--7) of the 18 strong-candidate optical (visible--near-infrared) counterparts to \Chandra-detected X-ray sources.
Examination of the near-infrared color--color diagram for all 2MASS sources within 6\arcmin\ of the pointing direction indicates those that are strong-candidate counterparts to X-ray sources are distributed as the field sources---i.e., as reddened main-sequence stars.
Although most Galactic-plane 2MASS objects are normal stars, the objects identified with the X-ray sources need not be normal stars:
For example, the X-ray emission may originate in an accreting compact companion.
Figure ~\ref{f:2MASS_colors} shows the X-ray hardness ratio of the 9 brighter X-ray sources versus the infrared color J-H of the corresponding strong 2MASS counterparts. 
The x-ray-soft and bluer infrared sources in the lower left corner of the figure are likely foreground stars.
The x-ray-hard and reddened sources in the upper right hand corner
may be X-ray binaries and/or AGN.

\subsubsection{WIYN observations\label{s:wiyn}}

The images we describe in Section~\ref{s:optical} allow us to measure or limit the optical magnitudes of several other X-rays sources.
A few sources were not covered in all sub-frames of the image stack and so were measured from individual exposures.  
Three additional optical counterparts to the X-ray sources were detected in the low exposure guide sector allowing improved frame registration. 
Table~\ref{t:omags} gives the detected magnitudes and upper limits.

\section{Further Discussion and Summary\label{s:summary}}

Using the \Chandra\ X-ray Observatory, we continued our search (Becker et al. 2004, Weisskopf et al. 2006) for possible X-ray counterparts to the intriguing \gray\ source now known as \FL.
We found 44 X-ray sources in a field centered on the \FL\ position, located along the line-of-sight toward the $\gamma$-Cyg SNR.
Only one of these sources, \#20, can reasonably and with high confidence be taken as the X-ray counterpart to the \gray\ source. 

There are a number of reasons supporting this conclusion.
First and foremost, our X-ray source \#20, is only 14.7\arcsec\ distant from the best-fitting \gray\ timing position. 
In addition, this separation is within the combined statistical and systematic errors on that position. 
There are also no other X-ray sources detected within 66\arcsec\ of the \gray\ source position to a {\sl Chandra} source-detection limit of $\sim$10$^{-15}$ erg~cm$^{-2}$~s$^{-1}$ in the 0.5$-$8.0 keV bandpass
making it highly unlikely that any of the other X-ray sources in the field are candidate counterparts.
Furthermore, the spectrum of source \#20 has a shape consistent with soft ($\log (T_{\infty}) \sim 6.0$ to $6.5$) thermal emission as expected from a young neutron star though perhaps somewhat higher than expected from the spin-down age of 77000~yr estimated for \FL.
There is also a hint of extended diffuse X-ray emission in the vicinity of source \#20 that 
 may be an associated PWN.

With source \#20 as the counterpart, we infer $F_\gamma/F_{\rm X} \sim 1.1 \times 10^4$, not atypical of young isolated neutron stars (e.g., Becker 2009). 
If source \#20 is not the X-ray counterpart, the flux ratio is at least 30$\times$ larger, 
which would be substantially larger
than the observed ratio for other \gray\ pulsars.

A similar argument using the optical data also supports source \#20 as the counterpart: our r$^\prime >25.2$ limit implies a lower limit of $F_{\rm X}/F_{\rm V} \approx 250$, with some uncertainty due to extinction. 
This is already larger than the maximum value for X-ray binaries ($\sim$15) or BL Lacs ($\sim$100) and  is approaching typical values for isolated neutron stars  ($\sim 10^{3-4}$) (e.g. Schwope et al. 1999). 
Thus, based on the X-ray/optical evidence alone, source \#20 is likely an isolated neutron star and is the likely counterpart for PSR J2021+4026.

Finally, the X-ray spectrum has a shape consistent with the soft thermal emission expected from a young neutron star. 
This emission likely represents a fraction of the stellar surface heated by back-flowing particles generated by magnetospheric activity (eg. Harding \& Muslimov 2001). 
At present the fitted parameters suggest that this thermal component implies a relatively large distance, $\approx 6$ kpc, incompatible with an association with SNR G78+2.1. 
However, the fits also indicate a complex spectrum with at least two components; much higher S/N data will be needed to extract strong spectral constraints. 
Of course, a heated polar cap suggests that sensitive observations should also be able to detect X-ray pulsations at the 265~ms spin period, the definitive test of the counterpart's association.

\acknowledgements

The work of MCW, DAS, RFE, SLO, and AFT is supported by the \Chandra\ Program.
The \Chandra\ data was obtained in response to proposal number 11500575
by the Chandra X-ray Observatory Center, which is operated by the Smithsonian Astrophysical Observatory for and on behalf of the National Aeronautics Space Administration under contract NAS8-03060. Support for this work was also provided to PSP in response to this proposal through Chandra Award Number GO0-11086A issued by the Chandra X-ray Observatory Center.
The work of RWR was supported in part by NASA grant NNX08AW30G.
The \textit{Fermi} LAT Collaboration acknowledges generous ongoing support
from a number of agencies and institutes that have supported both the
development and the operation of the LAT as well as scientific data analysis.
These include the National Aeronautics and Space Administration and the
Department of Energy in the United States, the Commissariat \`a l'Energie Atomique
and the Centre National de la Recherche Scientifique / Institut National de Physique
Nucl\'eaire et de Physique des Particules in France, the Agenzia Spaziale Italiana
and the Istituto Nazionale di Fisica Nucleare in Italy, the Ministry of Education,
Culture, Sports, Science and Technology (MEXT), High Energy Accelerator Research
Organization (KEK) and Japan Aerospace Exploration Agency (JAXA) in Japan, and
the K.~A.~Wallenberg Foundation, the Swedish Research Council and the
Swedish National Space Board in Sweden.
Additional support for science analysis during the operations phase is gratefully
acknowledged from the Istituto Nazionale di Astrofisica in Italy and the Centre National d'\'Etudes Spatiales in France.
Our analyses utilized software tools provided by the Chandra X-ray Center (CXC) in the application package CIAO and from the High-Energy Astrophysics Science Archive Research Center (HEASARC, operated by the NASA Goddard Space Flight Center, Greenbelt MD, and by the Smithsonian Astrophysical Observatory, Cambridge MA).
We also thank Mike Wolff for a careful critique of the manuscript.

\begin{deluxetable}{crrrrrrrr}
\tabletypesize{\tiny}
\tablewidth{0pc}
\tablecaption{\Chandra\ X-ray sources on ACIS-S3 in the Fermi-LAT PSR J021+4026 field. \label{t:data_x}}
\tablehead{
\colhead{Source} & \colhead{RA(J2000)} & \colhead{Dec(J2000)}
 & \colhead{$\theta_{\rm ext}$} & \colhead{$ C_{x}$} 
 & \colhead{$\sigma_{x}$} & \colhead{$\epsilon_{99}$}
 & \colhead{$(H-S)/T$} & \colhead{$M/T$} \\
\colhead{} & \colhead{$^{\rm h}\: \ ^{\rm m}\ \ ^{\rm s}\: $\ \ \ \ }  
 & \colhead{\ $\ \arcdeg\: \ \ \arcmin\: \ \ \arcsec$\ \ \ }   
 & \colhead{$\arcsec$} & \colhead{} & \colhead{$\arcsec$} 
 & \colhead{$\arcsec$}  
 & \colhead{}
 & \colhead{} \\ 
\colhead{(1)} & \colhead{(2)} & \colhead{(3)} & \colhead{(4)} & \colhead{(5)} 
 & \colhead{(6)} & \colhead{(7)}  & \colhead{(8)} & \colhead{(9)} \\
} 
\startdata
1  & 20 21  9.043 & 40 27  3.28 & 1.8 &  17 & 0.40 & 0.81 &  0.14 $\pm$ 0.20  &  0.66 $\pm$ 0.13  \\
2  & 20 21 10.472 & 40 28 44.47 & 3.3 &  12 & 0.64 & 1.28 &                 &                   \\
3  & 20 21 11.429 & 40 28  4.87 & 1.9 & 497 & 0.31 & 0.61 & -0.41 $\pm$ 0.03  &  0.47 $\pm$ 0.02  \\
4  & 20 21 12.731 & 40 28 31.91 & 1.4 &  14 & 0.38 & 0.76 &  0.10 $\pm$ 0.25  &  0.41 $\pm$ 0.10    \\
5  & 20 21 13.315 & 40 26  3.13 & 0.9 &  10 & 0.35 & 0.70 &                 & \\
6  & 20 21 13.576 & 40 25 55.80 & 1.4 &  25 & 0.34 & 0.69 &  0.18 $\pm$ 0.13  &  0.67 $\pm$ 0.09   \\
7  & 20 21 13.665 & 40 28 59.60 & 1.7 &   9 & 0.46 & 0.91 &                 & \\
8  & 20 21 14.337 & 40 25 20.37 & 0.9 &   7 & 0.36 & 0.72 &                 & \\
9  & 20 21 16.969 & 40 25 17.18 & 1.4 &  18 & 0.36 & 0.72 & -0.37 $\pm$ 0.23  &  0.13 $\pm$ 0.06    \\
10 & 20 21 19.524 & 40 25 32.86 & 1.4 &  12 & 0.39 & 0.78 &                 & \\
11 & 20 21 20.694 & 40 24  1.86 & 1.3 &  13 & 0.37 & 0.74 &                 & \\
12 & 20 21 21.132 & 40 27 46.35 & 0.8 &  11 & 0.34 & 0.68 &                 & \\
13 & 20 21 22.317 & 40 28 50.75 & 3.7 &  12 & 0.71 & 1.43 &                 & \\
14 & 20 21 25.164 & 40 28 13.34 & 1.3 &  33 & 0.33 & 0.67 & 0.90  $\pm$ 0.08  &  0.04 $\pm$ 0.04    \\
15 & 20 21 25.750 & 40 27 44.10 & 1.1 &   7 & 0.39 & 0.78 &                 & \\
16 & 20 21 26.087 & 40 23  5.22 & 1.0 &   7 & 0.38 & 0.76 &                 & \\
17 & 20 21 28.658 & 40 24 15.88 & 1.0 &   8 & 0.37 & 0.74 &                 & \\
18 & 20 21 29.773 & 40 24 55.09 & 0.8 &  11 & 0.33 & 0.67 &                 & \\
19 & 20 21 30.342 & 40 29 48.31 & 3.2 &  52 & 0.40 & 0.81 & 0.78  $\pm$ 0.10  &  0.18 $\pm$ 0.07  \\
20 & 20 21 30.733 & 40 26 46.04 & 1.3 & 281 & 0.31 & 0.61 & 0.02  $\pm$ 0.03  &  0.75 $\pm$ 0.02   \\
21 & 20 21 30.801 & 40 25 16.38 & 1.3 &  20 & 0.35 & 0.70 &  0.9  $\pm$ 0.13  &  0.09 $\pm$ 0.08     \\
22 & 20 21 31.385 & 40 22 56.47 & 1.4 &  26 & 0.34 & 0.69 & 0.14  $\pm$ 0.12  &  0.71 $\pm$ 0.09   \\
23 & 20 21 31.889 & 40 24 25.51 & 1.4 &  11 & 0.39 & 0.79 &                 & \\
24 & 20 21 32.659 & 40 28 21.38 & 2.2 &  11 & 0.50 & 1.01 &                 & \\
25 & 20 21 32.905 & 40 24 20.88 & 0.8 &  14 & 0.33 & 0.66 &                 & \\
26 & 20 21 33.031 & 40 23  0.62 & 2.3 &  11 & 0.52 & 1.05 &                 & \\
27 & 20 21 33.650 & 40 29  8.97 & 2.8 &  48 & 0.39 & 0.78 &  0.80 $\pm$ 0.11  &  0.25 $\pm$ 0.09  \\
28 & 20 21 34.097 & 40 25 26.50 & 1.3 &  12 & 0.38 & 0.76 &                 & \\
29 & 20 21 34.559 & 40 23 19.16 & 2.0 &  33 & 0.37 & 0.73 &  0.86 $\pm$ 0.10  &  0.09 $\pm$ 0.07     \\
30 & 20 21 35.268 & 40 28 35.85 & 2.0 &  16 & 0.43 & 0.87 &                 & \\
31 & 20 21 35.485 & 40 28 13.57 & 1.9 &   6 & 0.55 & 1.10 &                 & \\
32 & 20 21 37.579 & 40 29 58.09 & 3.6 & 248 & 0.33 & 0.67 & -0.21 $\pm$ 0.05  &  0.55 $\pm$ 0.03    \\
33 & 20 21 38.401 & 40 29 35.49 & 2.8 &  53 & 0.38 & 0.76 &  0.74 $\pm$ 0.11  &  0.22 $\pm$ 0.08  \\
34 & 20 21 38.431 & 40 24 42.77 & 1.7 &  94 & 0.32 & 0.64 & -0.45 $\pm$ 0.07  &  0.43 $\pm$ 0.05    \\
35 & 20 21 38.579 & 40 24 14.88 & 1.3 &   7 & 0.43 & 0.86 &                 & \\
36 & 20 21 39.214 & 40 27  9.90 & 1.0 &   9 & 0.36 & 0.73 &                 & \\
37 & 20 21 40.083 & 40 24  9.43 & 2.5 &  11 & 0.55 & 1.11 &                 & \\
38 & 20 21 43.107 & 40 23 53.76 & 2.6 &  48 & 0.38 & 0.76 & -0.08 $\pm$ 0.10  &  0.72 $\pm$ 0.07   \\
39 & 20 21 44.548 & 40 29 34.34 & 3.3 &  19 & 0.55 & 1.10 &                 & \\
40 & 20 21 47.294 & 40 24 54.84 & 3.2 &  28 & 0.47 & 0.95 &  0.99 $\pm$ 0.16  &  0.01 $\pm$ 0.12   \\
41 & 20 21 47.584 & 40 26 57.50 & 1.9 &  18 & 0.41 & 0.82 & -0.53 $\pm$ 0.44  &  0.20 $\pm$ 0.13    \\
42 & 20 21 52.529 & 40 25  7.72 & 4.2 &  66 & 0.44 & 0.87 & 0.79  $\pm$ 0.09  &  0.16 $\pm$ 0.07    \\
43 & 20 21 57.170 & 40 26 24.36 & 7.8 &  29 & 0.93 & 1.87 &  0.94 $\pm$ 0.23  &  0.15 $\pm$ 0.18    \\
44 & 20 21 57.751 & 40 26 46.78 & 5.8 & 105 & 0.46 & 0.92 & -0.42 $\pm$ 0.12  &  0.31 $\pm$ 0.04    \\
 \enddata
 \end{deluxetable}

\begin{deluxetable}{lccccc}
\tabletypesize{\scriptsize}
\rotate
  \tablewidth{0pc}
\tablecaption{Fits to single models with absorption.\label{t:smodels}}
  \tablehead{parameter       & {\tt powerlaw}             &{\tt bbodyrad} & {\tt nsa}$^a$            &{\tt nsmax-1260}$^b$& {\tt nsmax-130190}$^b$}
  \startdata
\nH                          & 1.58                       &0.79           & 1.03                     & 1.08               & 1.08            \\
$\sigma_{nH}$ $^c$           &(-0.20,+0.22)               &(-0.15,+0.18)  & (-0.19,+0.21)            &(-0.19,+0.21)       &(-0.18,+0.21)    \\
pl-index                     &  4.86                      &               &                          &                    &                 \\
$\sigma_{pl_index}$ $^c$     &(-0.48,+0.55)               &               &                          &                    &                 \\
$\log (T_{\infty}$              &     	                  &6.52$^d$       & 6.16                     & 6.12               & 6.10            \\
$\sigma_{\log (T_{\infty})}$ $^c$&                            &(-0.05,+0.04)  &(-0.08.+0.07)             &(-0.08,+0.08)       &(-0.08,+0.08)    \\
norm                         & 1.06 $\times10^{-4}$       & 0.80          & 1.8$\times10^{-9}$       & 0.56               & 0.83            \\
$\sigma_{norm}$ $^c$         &(-0.32,+0.51)$\times10^{-4}$&(-0.37,+0.83)  &(-1.2,+4.5)$\times10^{-9}$&(-0.38.+1.44)       &(-0.19,+0.21)    \\
D(kpc)$^e$                   &                            & 174.          & 23                       & 16                 & 13              \\
$\sigma_{D}$  $^c$           &                            & (-52,+63)     & (-4,+17)                 &(-8,+12)            &(-1.4,+1.8)      \\

R$_{em}$(km)$^f$             &                            & 0.1           & 0.9                      & 1.2                & 1.5             \\

\chisq                       & 23.3                       & 30.6          & 29.8                     & 28.5               & 27.7            \\
degrees of freedom                          & 22                         & 22            & 22                       & 22                 & 22              \\
Flux $\times10^{14}$ (\fcgs) & 43                         &4.9            &8.0                       &8.9                 &2.0              \\
 \enddata
\\[-4ex] 
 \tablecomments{\\
$^a$ The mass, radius, and magnetic field were fixed at $1.358M_\odot$, 12.996km, and $1.0\times10^{13}$ gauss respectively.\\
$^b$ The mass and radius of the neutron star were chosen as for the nsa model so that the input, the gravitational redshift 1+z, was fixed at 1.15.\\
$^c$ Uncertainties based on considering only 1 interesting parameter, i.e. the bounds indicated by the minimum \chisq$+1$.\\
$^d$ Assumes M/R=1.358/12.996 $M_\odot$/km.\\
$^e$ Derived from the different normalizations assuming a NS mass and radius $1.358M_\odot$, 12.996km.\\
$^f$ Rough estimate of the size of the emitting region by scaling the distance to 1.5 kpc.\\
}
 \end{deluxetable}

\clearpage
\begin{deluxetable}{lcccc}
\tabletypesize{\scriptsize}
  \tablewidth{0pc}
\tablecaption{Fits to dual models with absorption.\label{t:dmodels}}
  \tablehead{parameter &  {\tt bbodyrad} & {\tt nsa}$^a$ & {\tt nsmax-1260}$^b$ & {\tt nsmax-130190}$^b$}
  \startdata
\chisq                          & 16.38        & 16.24        & 16.36        &  16.38          \\
degrees of freedom                             & 20           & 20           & 20           &  20             \\
f statistic                     & 4.23         & 8.34         & 7.03         &  6.93           \\
probability                     & 0.029        & 0.0023       & 0.0049       &  0.0052         \\
\nH                             & 0.76         & 0.97         & 0.98         & 0.98            \\
$\sigma_{nH}$ $^e$              &(-0.16,+0.19) &(-0.18,+0.22) &(-0.19,+0.23) &(-0.18,+0.23)    \\
$\log (T_{\infty}$                 & 6.41         & 6.01         & 5.98         & 6.00         \\
$\sigma_{\log (T_{\infty}}$ $^e$   &(-0.08,+0.07) &(-0.12.+0.11) &(-0.13,+0.11) &(-0.14,+0.10) \\
D$^d$(kpc)                      & 91.          & 6.8          & 6.0          & 5.7             \\
$\sigma_{D}$  $^e$              &(-15,+39)     &(-4.7,+9.5)   &(-4.3,+9.4)   &(-4.4,+6.2)      \\
R$_{em}$(km)$^g$                & 0.2          & 2.8          & 3.3          & 3.4             \\
$\Gamma$		        & 1.2	       & 1.10         &0.73          & 0.73            \\
$\sigma_\Gamma$ $^e$            &(-1.5,+1.2)   &(-1.6,+1.3)   &(-2.10,+1.44) &(-2.2,+1,5)      \\
PL norm                         & 1.5          & 1.26         & 0.69         & 0.68            \\	
$\sigma_{norm}\times10^6$ $^e$  &(-1.5,+6.0)   &(-1.13,+5.88) &(-0.57,+1.05) &(-0.56,+7.02)    \\
Flux$^f$~$\times10^{14}$ (\fcgs)& 8.4          & 15.8         & 16.3         & 16.8            \\
	
 \enddata
\\[-4ex] 
 \tablecomments{\\
$^a$ The mass, radius, and magnetic field were fixed at $1.358\Msun$, 12.996km, and $1.0\times10^{13}$ gauss respectively.\\
$^b$ The mass and radius of the neutron star were chosen as for the nsa model so that the gravitational redshift 1+z was fixed at 1.15.\\
$^d$ Derived from the different normalizations assuming, where necessary, a NS mass and radius of $1.358\Msun$, 12.996km respectively.\\
$^e$ Uncertainties based on considering only 1 interesting parameter, i.e. the bounds indicated by the value of the parameter in question at the minimum \chisq$+1$.\\
$^f$ Unabsorbed flux. \\
$^g$ Rough estimate of the size of the emitting region by scaling the distance to 1.5 kpc.\\
}
 \end{deluxetable}

\begin{deluxetable}{ccccc}
\tabletypesize{\scriptsize}
  \tablewidth{0pc}
\tablecaption{Fit to a two-temperature mekal model for source \#3.\label{t:s3_spectrum}}
  \tablehead{\nH    & $kT_1$ & $kT_2$ & $norm_1$ & $norm_2$ \\
($10^{22} cm^{-3}$) & (keV)  & (keV)   & $\times 10^4$        & $\times 10^5$}
  \startdata
$0.44$  &  $0.20$  & $1.03$ &  $1.37 $   &  $2.36 $    \\
$(-0.13,+0.11)$ & $(-0.02,+0.04)$ & $(\pm 0.07)$ & $(-0.9.+1.9)$ & $(-0.32,+0.25)$\\
 \enddata
\\[-4ex] 
\tablecomments{\\
Uncertainties based on considering only 1 interesting parameter, i.e. the bounds indicated by the value of the parameter in question at the minimum \chisq$+1$.\\
}
 \end{deluxetable}

\clearpage

\begin{deluxetable}{cccccccccccc}
\tabletypesize{\scriptsize}
\rotate
\tablewidth{0pc}
\tablecaption{Candidate cataloged counterparts to X-ray sources in the \FL\ field.\label{t:data_oi}}
\tablehead{(1) & (2) & (3) & (4) & (5) & (6) & (7) & (8) & (9) & (10) & (11) & (12)}
\startdata
  { } & \multicolumn{6}{c}{USNO (optical) candidate counterpart} 
 & \multicolumn{5}{c} {2MASS (infrared) candidate counterpart} \\ 
  X-ray Source & RA(J2000) & Dec(J2000) & $\sigma_{o}^a$ & $\delta_{ox}$ & I
 & $p_{o}(\delta_{ox},{\rm I})$
 & RA(J2000)$^b$ & Dec(J2000)$^b$ & $\delta_{ix}$ & K$_{s}$ & $p_{i}(\delta_{ix},{\rm K}_{s})$ \\
  {} & \multicolumn{1}{c}{$^{\rm h}\: \ ^{\rm m}\ \ ^{\rm s}\: $\ \ \ \ } 
 & \multicolumn{1}{c}{\ $\ ^\circ\: \ \ \arcmin\: \ \ \arcsec$\ \ \ }
 & $\arcsec$ & $\arcsec$ & mag & \%
 & \multicolumn{1}{c}{$^{\rm h}\: \ ^{\rm m}\: \ \ ^{\rm s}\: $\ \ \ \ } 
 & \multicolumn{1}{c}{\ $\ \arcdeg\: \ \ \arcmin\ \ \arcsec$\ \ \ } 
 & $\arcsec$ & mag & \% \\ \hline\\[-2ex]
1  & 20 21 09.036 & 40 27 03.51 & (0.04,0.07) & 0.24 & 18.58 & 0.36 & 20 21 09.035 & 40 27 03.30 & 0.12 & 13.625 & 0.18 \\
3  & 20 21 11.433 & 40 28 05.01 & (0.04,0.10) & 0.18 & 13.15 & 0.07 & 20 21 11.424 & 40 28 04.71 & 0.18 & 11.155 & 0.01 \\
5  &              &             &             &      &       &      & 20 21 13.324 & 40 26 02.91 & 0.24 & 14.501 & 0.27 \\
6  &              &             &             &      &       &      & 20 21 13.596 & 40 25 55.80 & 0.24 & 14.175 & 0.21 \\
8  & 20 21 14.343 & 40 25 20.63 & (0.04,0.03) & 0.24 & 14.91 & 0.12 & 20 21 14.348 & 40 25 20.37 & 0.12 & 12.064 & 0.04 \\
9  & 20 21 06.967 & 40 25 17.44 & (0.04,0.04) & 0.24 & 13.71 & 0.10 & 20 21 16.967 & 40 25 17.17 & 0.06 & 12.559 & 0.06 \\
10 & 20 21 19.557 & 40 25 33.02 & (0.04,0.03) & 0.42 & 18.19 & 0.30 & 20 21 19.549 & 40 25 32.59 & 0.36 & 13.869 & 0.20 \\
11 & 20 21 20.702 & 40 24 02.14 & (0.04,0.07) & 0.30 & 14.58 & 0.12 & 20 21 20.700 & 40 24 01.97 & 0.12 & 12.249 & 0.05 \\
15 & 20 21 25.749 & 40 27 44.39 & (0.04,0.08) & 0.30 & 17.21 & 0.22 & 20 21 25.767 & 40 27 44.06 & 0.18 & 13.722 & 0.18 \\
17 & 20 21 28.657 & 40 24 15.97 & (0.04,0.02) & 0.12 & 19.07 & 0.35 & 20 21 28.679 & 40 24 15.50 & 0.42 & 15.083 & 0.44 \\
18 &              &             &             &      &       &      & 20 21 29.779 & 40 24 55.53 & 0.42 & 15.002 & 0.35 \\
22 & 20 21 31.425 & 40 22 56.71 & (0.04,0.16) & 0.54 & 17.47 & 0.18 & 20 21 31.422 & 40 22 56.41 & 0.42 & 13.188 & 0.09 \\
32 & 20 21 37.566 & 40 29 58.32 & (0.04,0.07) & 0.30 & 14.54 & 0.09 & 20 21 37.575 & 40 29 57.90 & 0.18 & 13.302 & 0.09 \\
34 & 20 21 38.410 & 40 24 42.97 & (0.04,0.08) & 0.30 & 13.11 & 0.08 & 20 21 38.427 & 40 24 42.63 & 0.12 & 11.476 & 0.02 \\
37 &              &             &             &      &       &      & 20 21 40.056 & 40 24 09.63 & 0.36 & 14.465 & 0.65 \\
38 & 20 21 43.117 & 40 23 54.22 & (0.04,0.08) & 0.48 & 13.68 & 0.11 & 20 21 43.116 & 40 23 54.05 & 0.30 & 11.967 & 0.04 \\
39 & 20 21 44.515 & 40 29 34.03 & (0.04,0.03) & 0.48 & 18.00 & 0.55 & 20 21 44.538 & 40 29 33.44 & 0.90 & 14.003 & 0.45 \\
41 & 20 21 47.628 & 40 26 57.94 & (0.04,0.32) & 0.66 & 17.33 & 0.24 &              &             &      &        &      \\
42 &              &             &             &      &       &      & 20 21 52.555 & 40 25 08.16 & 0.54 & 13.863 & 0.25 \\
43 &              &             &             &      &       &      & 20 21 57.240 & 40 26 23.10 & 1.50 & 14.723 & 2.28 \\
\enddata 
\tablecomments{
$^a$ USNO RMS positional uncertainty in each axis (RA, Dec) \\
$^b$ 2MASS RMS positional uncertainty $\sigma_{i}=0\farcs08$ per axis \\
}
\end{deluxetable}

\clearpage

\begin{deluxetable}{ccccc}
\tabletypesize{\scriptsize}
\tablewidth{0pc}
\tablecaption{2MASS infrared photometry of strong-candidate counterparts to X-ray sources. \label{t:2MASS_colors}}
\tablehead{(1) & (2) & (3) & (4) }
\startdata
X-ray Source   & J & J$-$H & H$-$K$_s$ \\ \hline\\
1 & $ 15.583 \pm 0.043 $  & $ 1.436 \pm 0.068 $ & $ 0.522 \pm 0.073 $  \\ 
3 & $ 11.927 \pm 0.018 $  & $ 0.675 \pm 0.028 $ & $ 0.097 \pm 0.027 $  \\ 
5 & $ 16.031 \pm 0.068 $  & $ 1.285 \pm 0.107 $ & $ 0.245 \pm 0.133 $  \\ 
6 & $ 15.772 \pm 0.056 $  & $ 1.195 \pm 0.087 $ & $ 0.402 \pm 0.099 $  \\ 
8 & $ 13.113 \pm 0.021 $  & $ 0.698 \pm 0.030 $ & $ 0.351 \pm 0.030 $  \\ 
9 & $ 13.045 \pm 0.021 $  & $ 0.387 \pm 0.030 $ & $ 0.099 \pm 0.032 $  \\ 
10 & $ 15.278 \pm 0.040 $ & $ 1.177 \pm 0.059 $ & $ 0.232 \pm 0.073 $  \\ 
11 & $ 13.154 \pm 0.022 $ & $ 0.713 \pm 0.033 $ & $ 0.192 \pm 0.034 $  \\ 
15 & $ 15.133 \pm 0.038 $ & $ 0.991 \pm 0.064 $ & $ 0.420 \pm 0.074 $  \\ 
17 & $ 16.021 \pm 0.089 $ & $ 0.945 \pm 0.121 $ & $-0.007 \pm 0.190 $  \\ 
18 & $ 16.046 \pm 0.136 $ & $ 0.442 \pm 0.164 $ & $ 0.602 \pm 0.180 $  \\ 
22 & $ 14.761 \pm 0.027 $ & $ 1.238 \pm 0.041 $ & $ 0.335 \pm 0.047 $  \\ 
32 & $ 13.887 \pm 0.021 $ & $ 0.534 \pm 0.031 $ & $ 0.051 \pm 0.043 $  \\ 
34 & $ 12.045 \pm 0.021 $ & $ 0.517 \pm 0.030 $ & $ 0.052 \pm 0.028 $  \\
37 & $ 15.822 \pm 0.064 $ & $ 1.119 \pm 0.093 $ & $ 0.238 \pm 0.124 $  \\
38 & $ 12.582 \pm 0.021 $ & $ 0.480 \pm 0.030 $ & $ 0.135 \pm 0.028 $  \\
39 & $ 15.477 \pm 0.108 $ & $ 0.974 \pm 0.129 $ & $ 0.500 \pm 0.123 $  \\ 
42 & $ 15.995 \pm 0.062 $ & $ 1.307 \pm 0.097 $ & $ 0.825 \pm 0.101 $  \\ 
43 & $ 16.249 \pm 0.075 $ & $ 1.280 \pm 0.128 $ & $ 0.246 \pm 0.160 $  \\ 
 \enddata
 \end{deluxetable}

\begin{deluxetable}{rrrrrrrrrrrr}
\tabletypesize{\scriptsize}
\tablecaption{\label{t:omags} WIYN Optical Magnitudes}
\tablehead{
\colhead{Source} & \colhead{$i^\prime$} & \colhead{$r^\prime$} & \colhead{Source} & \colhead{$i^\prime$} & \colhead{$r^\prime$} & \colhead{Source} & \colhead{$i^\prime$} & \colhead{$r^\prime$} & \colhead{Source} & \colhead{$i^\prime$} & \colhead{$r^\prime$} 
}
\startdata  
1  & 18.89  & 21.04  & 2  &$\sim$24.4 & $>^a$  & 3  & 15.52   & 15.77  & 4  & 17.96 & 19.95 \\
5  & 19.30 & 21.36 & 6  & 19.02 & 19.02 & 7  & $>$    &  $>$  & 8  & 15.63 & 17.07 \\ 
9  & 15.15 & 15.50  & 10 & 18.39 & 20.39  & 11 & 15.63 & 15.75  & 12 & $>$  & $>$ \\
13 & 22.40 & $>$  & 14 & $>$    & $>$  & 15 & 17.94 & 19.82 & 16 & $>$    & $>$  \\
17 & 19.37 & 21.83 & 18 & 18.13 & 20.15 & 19 &\nodata &\nodata & 21 & $>$    & $>$   \\
22 & 17.82 & 19.94 & 23 & 20.92 & 23.29 & 24 &$\sim24.4$&$\sim25.5$ & 25 & $>$    & $>$  \\
26 & $>$    & $>$  & 27 & 22.45 & \nodata & 28 & \nodata& \nodata & 29 & $>$    & $>$   \\
\enddata
\tablenotetext{a}{Magnitude limits: $>$ = $>24.2$ ($i^\prime$), = $>25.3$ ($r^\prime$)}
\end{deluxetable}
\bigskip

\clearpage

\clearpage

\clearpage

\clearpage

\clearpage

\clearpage

\clearpage

\clearpage

\begin{thebibliography}{47}
%
\bibitem[Abdo et al.]{abd08} Abdo, A.~A. et al. 2008, Science, 322, 1218
%
\bibitem[Abdo et al. (2009a)]{abd09a} Abdo, A.~A. et al. 2009a, ApJS, 183, 46
%
\bibitem[Abdo et al. (2009b)]{abd09b} Abdo, A.~A. et al. 2009b, Science, 325, 840 (0FGL)
%
\bibitem[Abdo et al. (2010a)]{abd10a} Abdo, A.~A. et al. 2010a, ApJS, 
187, 460
%
\bibitem[Abdo et al. (2010b)]{abd10b} Abdo, A.~A. et al. 2010b, ApJS, 
188, 405 (1FGL)
%
\bibitem[Abdo et al.(2010c)]{2010ApJ...711...64A} Abdo, A.~A., et al.\ 2010c, ApJ, 711, 64
%
\bibitem[Abdo et al. (2011)]{abd11} Abdo, A.~A. et al. 2011, in preparation for ApJS (2FGL)
%
\bibitem[Arnaud(1996)]{arn96} Arnaud, K.~A. 1996, in ASP Conf.~Ser.~101, Astronomical Data Analysis software Systems V, ed. ~G.~H. Jacoby \& J. Barnes (San Francisco: ASP), 17
%
\bibitem[Arons(1983)]{arons83} Arons, J. 1983, ApJ, 266, 215
%
\bibitem [Arons \& Scharleman (1979)]{as79} Arons, J. \& Scharlemann, E.~T. 1979, ApJ, 231, 854
%
\bibitem[Atwood et al.(2009)]{atwood2009} Atwood, W.~B. et al.\ 2009, ApJ, 697, 1071
%
\bibitem[Becker(2009)]{2009ASSL..357...91B} Becker, W.\ 2009, Astrophysics 
and Space Science Library, 357, 91 
%
\bibitem[Becker et al.(2004)]{bec04} Becker, W., Weisskopf, M.C., Arzoumanian, Z., Lorimer, D., Camilo, F., Elsner, R.F., Kanback, G., Reimer, O. \& Swartz, D.A. 2004, ApJ, 615, 897
%
\bibitem[Brazier et al.(1996)]{bra96} Brazier, K.~T.~S., Kanbach, G., Carrami\~{n}ana, A., Guichard, J. \& Merck, M. 1996, MNRAS, 281, 1033
%
\bibitem[Camilo et al.(2009)]{2009ApJ...705....1C} Camilo, F., et al.\ 2009, ApJ, 705, 1 
%
\bibitem [Caraveo et al.]{car10} Caraveo, P. et al. 2010,  ApJ, 725, L6
%
\bibitem [Cheng \& Zhang (1998)] {che98} Cheng, K. S. \& Zhang, L. 1998, ApJ, 498, 327
%
\bibitem [Cheng, Ho \& Ruderman (1986a)]{chr86a} Cheng, K. S., Ho, C. \& Ruderman, M.1986, ApJ, 300, 500
%
\bibitem [Cheng, Ho \& Ruderman (1986b)]{chr86b} Cheng, K. S., Ho, C. \& Ruderman, M.1986, ApJ, 300, 522
%
\bibitem[Daugherty \& Harding]{dau96} Daugherty, J. K. \& Harding, A. K. 1996, ApJ, 458, 278
%
\bibitem[De Luca et al.(2005)]{del05} De~Luca, A., Caraveo, 
P.~A., Mereghetti, S., Negroni, M., \& Bignami, G.~F.\ 2005, \apj, 623, 1051 
%
\bibitem[Dermer \& Sturner(1994)]{der94} Dermer, C.~D. \& Sturner, S.~J. 1994, ApJ, 420, L75
%
\bibitem[Dickey \& Lockman(1990)]{dic90} Dickey, J.~M. \& Lockman, F.~J. 1990, ARA\&A, 28, 215
%
%
\bibitem[Green(2009)]{gre09} Green, A.~J. 2009, Bulletin of the Astronomical Society of India, 37, 45
%
\bibitem[Halpern \& Holt(1992)]{hal92} Halpern, J.~P. \& Holt, S.~S. 1992, Nature, 357, 222
%
\bibitem[Harding et al.]{har08} Harding, A. K., Stern, J. V., Dyks, J. \& Frackowiak, M. 2008, ApJ, 680, 1378
%
\bibitem[Harding Muslimov]{har01} Harding, A. K. \& Muslimov, A. G. 2001, ApJ, 556, 987.
%
\bibitem[Harding Muslimov]{har05} Harding, A. K. \& Muslimov, A. G. 2005, Astr and Sp Science, 297, 63
%
\bibitem[Hartman et al.(1999)]{har99} Hartman, R.~C. et al. 1999, ApJS, 123, 79
%
\bibitem[Higgs, Landecker, \& Roger (1977)]{hig77} Higgs, L.~A., Landecker, T.~L. \& Roger, R.~S. 1977, AJ, 82, 718
%
\bibitem[Ho, Potekhin \& Chabrier 2008]{ho08} Ho, W. C. G., Potekhin, A. Y. \& Chabrier, G. 2008, ApJS, 178, 102
%
\bibitem[Hobbs, Edwards \& Manchester (2004)]{hob04} Hobbs, G.~B., Edwards, R.~T. \& Manchester, R.~N.\ 2004, MNRAS, 353, 1311
%
\bibitem[Hobbs, Edwards \& Manchester (2006)]{hob06} Hobbs, G.~B., Edwards, R.~T. \&  Manchester, R.~N.\ 2006, MNRAS, 369, 655
%
\bibitem[Jackson\& Halpern (2005) ]{jh05} Jackson, M. S. \& Halpern, J.P. 2005, ApJ, 633, 1114
%
\bibitem[Landecker, Roger \&Higgs (1980)]{lan80} Landecker, T.~L.,  Roger, R.~S. \& Higgs, L.~A. 1980, A\&AS, 39, 133
%
\bibitem[Mavromatakis(2003)]{mav03}Mavromatakis, F. 2003, A\&A 408, 237
%
\bibitem[Michel(1991)]{1991tnsm.book.....M} Michel, F.~C.\ 1991,``Theory of neutron star magnetospheres,'' (Chicago, UCP)  
%
\bibitem[Monet et al.(2003)]{mon03} Monet, D. et al. 2003, AJ, 125, 984
%
\bibitem[Muslimov \& Harding]{mus04} Muslimov, A. G. \& Harding, A. K. 2004, ApJ, 606, 1143
%
\bibitem[Pavlov et al. 1995]{pav95} Pavlov, G. G., Shibanov, Yu. A., Zavlin, V. E. \& Meyer, R. D. 1995, in “The Lives of the Neutron Stars,” ed. M.A. Alpar, U. Kiziloglu, \& J. van Paradijs (NATO ASI Ser. C, 450; (Dordrecht: Kluwer), 71 
%
\bibitem[Pavlov et al.(2010)]{2010ApJ...715...66P} Pavlov, G.~G., 
Bhattacharyya, S., \& Zavlin, V.~E.\ 2010, \apj, 715, 66 
%
\bibitem[Ransom, Eikenberry \& Middleditch (2002)]{ran02} Ransom, S.~M., Eikenberry, S.~S. \& Middleditch, J.\ 2002, AJ, 124, 1788 
%
\bibitem[Ray et al.(2011)]{ray11} Ray, P.~S. et al.\ 2011, ApJS, 194, 17
%
\bibitem[Romani \& Yadigaroglu(1995)]{rom95} Romani, R.~W. \& Yadigaroglu, I.~A. 1995, ApJ, 438, 314
%
\bibitem [Romani]{rom96} Romani, R. 1996, ApJ. 470, 469 
%
\bibitem [Ruderman \& Sutherland (1975)]{rus75} Ruderman, M. \& Sutherland, P. 1975, ApJ, 196, 51
%
\bibitem[Saz Parkinson et al. (2010)]{saz10} Saz Parkinson, P.~M. et 
al. 2010, ApJ, 725, 571
%
\bibitem[Schwope et al. (1999)]{sch99} Schwope, A. D. et al. 1999, A\&A, 341, L51
%
\bibitem[Skrutskie et al.(2006)]{skr06} Skrutskie, M. F. et al. 2006, AJ, 131, 1163
%
\bibitem[Smith et al. (2002)]{smi02} Smith, J. A. et al. 2002, AJ, 123, 2121
%
\bibitem[Sturner \& Dermer(1995)]{stu95} Sturner, S.~J. \& Dermer, C.~D. 1995, A\&A, 293, L17
%
\bibitem[Swanenburg,et al.(1981)]{swa81} Swanenburg, B.~N. et al. 1981, ApJ, 243, L69
%
\bibitem[Takata \& Chang]{tak07}  Takata, J. \& Chang, H.-K. 2007, ApJ, 670, 677
%
\bibitem[Tennant (2006)]{ten06} Tennant, A.~F. 2006, AJ, 132, 1372
%
\bibitem[Thompson (2001)]{tho01}  Thompson, D. J. 2001, HIGH ENERGY GAMMA-RAY ASTRONOMY: International Symposium. AIP Conference Proceedings, 558, 103
%
\bibitem[Tonry et al.(2004)]{tet04} Tonry, J. et al. 2004, ASSL, 300, 385
%
\bibitem[Trepl et al. (2010)]{tre10} Trepl, L. et al. 2010, MNRAS, 405, 339
%
\bibitem[Verner et al.(1993)]{ver93} Verner, D.~A., Yakovlev, D.~G., Band, I.~M. \& Trzhaskovskaya, M.~B. 1993, Atomic Data \& Nuclear Data Tables, 55, 233
%
\bibitem[Weisskopf et al. (2006)]{wei06} Weisskopf, M. C., Swartz, D. A., Carramiñana, A., Carrasco, L., Kaplan, D. L., Becker, W., Elsner, R., Kanbach, K., O’Dell, S. L. \& Tennant, A. F. 2006, ApJ, 652, 387
%
\bibitem[Wendker, Higgs \& Landecker (1991)]{wenk91}  Wendker, H. J., Higgs, L. A. \& Landecker, T. L. 1991, A\&A, 241, 551
%
\bibitem[Wilms, Allen, \& McCray(2000)]{wil00} Wilms, J., Allen, A., \& McCray, R. 2000, ApJ, 542, 914
%
\bibitem[Yadigaroglu \& Romani (1995)]{yad95} Yadigaroglu, I.-A. \& Romani, R. W. 1995, ApJ, 449, 211
%
\bibitem[Yakovlev et al. 2010]{yak10} Yakovlev, D. G. et al. 2010, MNRAS, 411, 1977
%
\end{thebibliography}
\end{document}